\documentclass[camera]{jpaper}
\usepackage[sort]{cite}

\usepackage{soul}

\usepackage{amsmath}
\usepackage[linesnumbered,ruled]{algorithm2e}
\usepackage{siunitx}
\usepackage{algpseudocode}
\usepackage{titlesec}
\usepackage[flushleft]{threeparttable}
\algrenewcommand\algorithmiccomment[2][\normalsize]{{#1\hfill\(\triangleright\) #2}}
\usepackage{geometry} \titlespacing*{\section}{0pt}{1pt}{0pt}
\titlespacing*{\subsection}{0pt}{1pt}{0pt}
\titlespacing*{\subsubsection}{0pt}{1pt}{3pt}

\usepackage{array}
\makeatletter
\let\MYcaption\@makecaption
\makeatother

\usepackage[font=footnotesize]{subcaption}

\makeatletter
\let\@makecaption\MYcaption
\makeatother

\usepackage{fixltx2e}

\usepackage{stfloats}
\usepackage[nolessnomore, italic]{mathastext}
\usepackage[T1]{fontenc}
\usepackage[usenames,dvipsnames,svgnames,table]{xcolor}
\usepackage{multirow}
\usepackage{hhline}
\usepackage[normalem]{ulem}
\usepackage{setspace}
\usepackage{indentfirst}
\usepackage[hang,flushmargin]{footmisc}

\usepackage{comment}
\usepackage{amssymb} 
\usepackage{xcolor}
\usepackage{textcomp}
\usepackage{fancyhdr}
\usepackage{microtype}
\usepackage{url}
\usepackage{flushend}
\usepackage{ulem}
\usepackage[bookmarks=true,breaklinks=true,letterpaper=true,colorlinks,linkcolor=black,citecolor=blue,urlcolor=black]{hyperref}
\usepackage{setspace}
\usepackage{mdframed} 
\usepackage{mathptmx} 
\usepackage{enumitem}
\usepackage{tikz}
\usepackage{subcaption} 
\usepackage{booktabs} 
\usepackage{caption}
\usepackage{multirow} 
\usepackage{float}
\usepackage{color}

\definecolor{blue(pigment)}{rgb}{0.2, 0.2, 0.7}
\definecolor{dgreen}{rgb}{0.0, 0.8, 0.0}
\usepackage{xcolor}
\definecolor{ocre}{RGB}{243,102,25}
\newmdenv[%
linecolor=blue(pigment),
backgroundcolor=blue(pigment)!10,
linewidth=0pt]{mytablebox}

\definecolor{amber}{rgb}{1.0, 0.49, 0.0}
\definecolor{darkgreen}{rgb}{0.0, 0.2, 0.13}
\definecolor{darkbyzantium}{rgb}{0.36, 0.22, 0.33}
\definecolor{darkseagreen}{rgb}{0.56, 0.74, 0.56}
\definecolor{darkspringgreen}{rgb}{0.09, 0.45, 0.27}
\definecolor{dollarbill}{rgb}{0.52, 0.73, 0.4}
\definecolor{MidnightBlue}{RGB}{46,46,208}

\newcommand\camone[1]{\noindent{\color{black}{#1}}} 
\newcommand\camtwo[1]{\noindent{\color{black}{#1}}} 
\newcommand\camthree[1]{\noindent{\color{black}{#1}}} 
\newcommand\camfour[1]{{\color{black}{#1}}} 
\newcommand\camfive[1]{{\color{black}{#1}}} 
\newcommand\camsix[1]{{\color{black}{#1}}} 
\newcommand\camseven[1]{{\color{black}{#1}}} 
\newcommand\camtracy[1]{{\color{black}{#1}}} 

\newcommand\cameight[1]{\noindent{\color{black}{#1}}}
\newcommand\camnine[1]{\noindent{\color{black}{#1}}}

\newcommand\cgiannou[1]{\noindent{\color{teal}}} 

\newcommand{\myName}{\emph{SynCron}}
\newcommand{\myEngine}{Synchronization Engine}
\newcommand{\myEngineShort}{SE}

\newcommand{\masterSE}{\emph{Master SE}}

\newcommand{\myTableShort}{ST}
\newcommand{\mySyncVar}{syncronVar}
\newcommand{\hier}{\emph{Hier}}
\newcommand{\naive}{\emph{Central}}
\newcommand{\ideal}{\emph{Ideal}}
\newcommand{\mpsync}{message-passing}
\newcommand{\Mpsync}{Message-passing}

\newcommand*\circled[1]{\tikz[baseline=(char.base)]{\node[shape=circle,fill,inner sep=0.5pt] (char) {\textbf{\textcolor{white}{#1}}};}}
\newcommand*\rectangled[1]{\tikz[baseline=(char.base)]{\node[shape=rectangle,fill,inner sep=1pt, minimum width=0.34cm, rounded corners] (char) {\textbf{\textcolor{white}{#1}}};}}

\newcommand{\affilNTUA}[0]{\textsuperscript{$\dagger$}}
\newcommand{\affilUoT}[0]{\textsuperscript{\textasteriskcentered}}
\newcommand{\affilETH}[0]{\textsuperscript{$\ddagger$}}
\newcommand{\affilUMA}[0]{\textsuperscript{\S}}

\usepackage{pifont}

\newcommand{\squeezeme}{ \setlength{\itemsep}{0pt}
     \setlength{\parsep}{1pt}
     \setlength{\topsep}{1pt}
     \setlength{\partopsep}{0pt}
     \setlength{\leftmargin}{1.5em}
     \setlength{\labelwidth}{1em}
     \setlength{\labelsep}{0.5em} }

\newcommand{\versionnum}[0]{5.3~---~\today~}

\begin{document}
\bstctlcite{IEEEexample:BSTcontrol} 

\setstretch{0.82}
\title{\textbf{\myName}: Efficient Synchronization Support \\ for Near-Data-Processing Architectures\vspace{-16pt}} 
\setstretch{0.84}

%


\author{
\hspace{-12pt}
\fontsize{11.4}{8}\selectfont
\parbox[t]{1.02\textwidth}{
{Christina Giannoula\affilNTUA\affilETH}\hspace{6pt}
{Nandita Vijaykumar\affilUoT\affilETH}\hspace{6pt}
{Nikela Papadopoulou\affilNTUA}\hspace{6pt}
{Vasileios Karakostas\affilNTUA}\hspace{6pt}
{Ivan Fernandez\affilUMA\affilETH}
{\vspace{-3pt}\textcolor{white}{}}
\\
{\vspace{3pt}\textcolor{white}{s}}\hspace{38pt}
{Juan Gómez-Luna\affilETH}\hspace{10pt}
{Lois Orosa\affilETH}\hspace{10pt} 
{Nectarios Koziris\affilNTUA}\hspace{10pt}%
{Georgios Goumas\affilNTUA}\hspace{10pt}%
{Onur Mutlu\affilETH} 
\\
\vspace{-7pt}
\emph{{\affilNTUA National Technical University of Athens\hspace{15pt} \affilETH ETH Z{\"u}rich  \hspace{15pt} \affilUoT University of Toronto \hspace{15pt} \affilUMA University of Malaga 
}}%
}
\vspace{-4pt}
}

\maketitle

\thispagestyle{plain} 
\pagestyle{plain}

\renewcommand\footnotelayout{\setstretch{0.82}}%
\setlength{\footnotemargin}{0in}

\setstretch{0.835}
\setlength{\intextsep}{2pt} 
\begin{abstract}

Near-Data-Processing (NDP) architectures present a promising way to alleviate data movement \camone{costs} and can provide significant performance and energy benefits to parallel applications. Typically, NDP architectures support several NDP units, each \camtwo{including} \camseven{multiple} simple cores placed close to memory. To fully leverage the benefits of NDP and achieve high performance for parallel workloads, efficient synchronization among the NDP cores of a system is necessary. However, supporting synchronization in many \camone{NDP} systems is challenging because they lack shared caches and hardware cache coherence support, which are commonly used for synchronization in multicore systems, and communication across \camseven{different} NDP units \camone{can be} expensive.

This paper comprehensively examines the synchronization problem in \camone{NDP} systems, and proposes \camnine{SynCron}, an end-to-end synchronization solution for NDP systems. \camnine{SynCron} adds \camseven{low-cost} hardware support \camone{near memory} for synchronization acceleration, \camseven{and avoids} the need for \camseven{hardware} cache coherence \camseven{support}. \camnine{SynCron} \camone{has three components\camtwo{:} 1)} a specialized cache memory structure to avoid memory accesses for synchronization and minimize latency \camone{overheads}, \camone{2)} a hierarchical message-passing communication protocol to minimize expensive communication across NDP units of the system, \camone{and 3)} a hardware-only overflow \camone{management} scheme to \camtwo{avoid} performance degradation \camone{when hardware resources \camtwo{for synchronization tracking} are exceeded.}

We evaluate \camnine{SynCron} \camone{using} a variety of parallel workloads, covering various contention scenarios. \camnine{SynCron} improves performance by 1.27$\times$ on average (up to 1.78$\times$) under \camone{high-contention} \camseven{scenarios,} and by \camone{1.35$\times$} on average (up to 2.29$\times$) under \camone{low-contention} real applications, \camseven{compared to state-of-the-art} approaches. \camnine{SynCron} reduces \camseven{system} energy consumption by 2.08$\times$ on average \camtwo{(up to 4.25$\times$)}. 

\end{abstract}
\section{Introduction}
\label{Introductionbl}

Recent advances in 3D-stacked memories~\cite{HBM,HMC,kim2015ramulator,HMC_old,HBM_old,Lee2016Simultaneous} have renewed interest in Near-Data Processing (NDP)~\cite{Mutlu2020AMP,Ahn2015Scalable,Ahn2015PIMenabled,Balasubramonian2014Near}. NDP involves performing computation close to where the application data resides. This alleviates the expensive data movement between processors and memory, \camtwo{yielding} significant performance improvements and energy savings in parallel applications. Placing low-power cores or special-purpose accelerators (hereafter \camone{called} NDP cores) close to the memory dies of high-bandwidth \camone{3D-}stacked memories is a commonly-proposed design for NDP systems~\cite{Mutlu2020AMP,Mutlu2019Processing,Ghose2019Workload,Ahn2015PIMenabled,Nair2015Active,Ahn2015Scalable,Hsieh2016accelerating,Pugsley2014NDC,Boroumand2018Google,Gokhale2015Near,Gao2016HRL,Hsieh2016TOM,Drumond2017mondrian,Liu2018Processing,Boroumand2019Conda,Gao2015Practical,Gao2017Tetris,Kim2016Neurocube,Kim2017GrimFilter,Survive2016Survive,Nai2017GraphPIM,Youwei2019GraphQ,Pattnaik2016Scheduling,fernandez2020natsa,Singh2019NAPEL,Singh2020NERO,Cali2020GenASM,Zhang2018GraphP,Kim2013memory,Tsai2018Adaptive,boroumand2017lazypim}. Typical NDP architectures support several NDP units connected to each other, with each unit comprising multiple NDP cores close to memory~\cite{Kim2013memory,Youwei2019GraphQ,Tsai2018Adaptive,Ahn2015Scalable,Hsieh2016TOM,Boroumand2018Google,Zhang2018GraphP}. Therefore, NDP architectures provide high levels of parallelism, low \camone{memory} access latency, and large aggregate memory bandwidth.

Recent \camtwo{research demonstrates} the benefits of NDP for parallel applications\camtwo{, e.g., for} \camone{genome analysis}~\cite{Kim2017GrimFilter,Cali2020GenASM}, graph processing~\cite{Ahn2015Scalable,Nai2017GraphPIM,Zhang2018GraphP,Youwei2019GraphQ,Ahn2015PIMenabled,Boroumand2019Conda,boroumand2017lazypim}, databases~\cite{Drumond2017mondrian,Boroumand2019Conda}, security~\cite{Gu2016Leveraging},  pointer-chasing \camtwo{workloads}~\cite{liu2017concurrent,choe2019concurrent,Hsieh2016accelerating,hashemi2016accelerating}, and neural networks~\cite{Boroumand2018Google,Gao2017Tetris,Kim2016Neurocube,Liu2018Processing}. In general, these applications exhibit high parallelism, low operational intensity, and \camone{relatively low} cache locality~\cite{Gagandeep2019Near,Awan2015Performance,Awan2016Node,oliveira2021pimbench,gomezluna2021upmem}, which make them suitable for NDP.

Prior works discuss the need for efficient synchronization primitives in NDP systems, such as locks~\cite{liu2017concurrent,choe2019concurrent} and barriers~\cite{Ahn2015Scalable,Gao2015Practical,Youwei2019GraphQ,Zhang2018GraphP}. Synchronization primitives are widely used by multithreaded applications~\cite{LeBeane2015Data,Tallent2010Analyzing,Strati2019AnAdaptive,Giannoula2018Combining,Elafrou2019Conflict,Suleman2009Accelerating,Joao2012Bottleneck,Joao2013Utility,Ebrahimi2011Parallel,Suleman2010Data}, and must be carefully designed to fit the underlying hardware requirements to achieve high performance. Therefore, to fully leverage the benefits of NDP for parallel applications, an effective synchronization solution for \camtwo{NDP} systems is necessary.

Approaches to support synchronization are typically of two types~\cite{herlihy2008art,Hoefler2004survey}.
First, synchronization primitives can be built through \emph{shared memory}, most commonly using the atomic read-modify-write (\emph{rmw}) operations provided by hardware. In CPU systems, atomic \emph{rmw} operations are typically implemented upon the underlying hardware cache coherence protocols, but \camone{many} NDP systems do \emph{not} support \camone{hardware} cache coherence (\camone{e.g.,}~\cite{Tsai2018Adaptive,Ghose2019Workload,Ahn2015Scalable,Zhang2018GraphP,Youwei2019GraphQ}). In GPUs and Massively Parallel Processing systems (MPPs), atomic \emph{rmw} operations can be implemented \camone{in} dedicated hardware atomic units, known as \emph{remote atomics}. However, synchronization \camone{using} remote atomics has been shown to be inefficient, since sending every update to a fixed location creates high global traffic and hotspots~\cite{Wang2019Fast,li2015fine,yilmazer2013hql,eltantawy2018warp,Mukkara2019PHI}. Second, synchronization can be implemented via a \emph{\mpsync{}} scheme, where cores exchange messages to reach an agreement. \camone{Some} recent NDP works (\camone{e.g.,}~\cite{Ahn2015Scalable,Gao2015Practical,Youwei2019GraphQ,Gu2020IPIM}) propose \mpsync{} barrier primitives among NDP cores of the system. However, these synchronization schemes are still inefficient, as we demonstrate in Section~\ref{Evaluationbl}\camtwo{, and also lack support for lock, semaphore and condition variable synchronization primitives.}

Hardware synchronization \camone{techniques} that do not rely on hardware coherence protocols and atomic \emph{rmw} operations have been proposed for multicore systems~\cite{abell2011glocks,abellan2010g,oh2011tlsync,Zhu2007SSB,Vallejo2010Architectural,Liang2015MISAR,Leiserson1992CM5,Sergi2016WiSync}. However, such synchronization schemes are tailored for the specific architecture of each system, and are not efficient or suitable for NDP systems (Section~\ref{Relatedbl}). For instance, CM5~\cite{Leiserson1992CM5} provides a barrier primitive \camone{via} a dedicated physical network, which would incur high hardware cost to be supported in \camone{large-scale} NDP systems. LCU~\cite{Vallejo2010Architectural} \camone{adds} a control unit to \emph{each} CPU core and a buffer to each memory controller, which would also incur high cost to implement in \emph{\camtwo{area-constrained}} NDP cores \camone{and controllers}. SSB~\cite{Zhu2007SSB} includes a small buffer attached to each controller of the last level cache (LLC) and MiSAR~\cite{Liang2015MISAR} introduces an accelerator distributed at the LLC. Both schemes are built on the shared cache \camone{level in} CPU systems, which most NDP systems do \emph{not} have. \camone{Moreover, in NDP systems \camone{with \emph{non-uniform} memory access times}, most of these prior schemes would incur significant performance overheads under high-contention scenarios. This is because they are oblivious to \camthree{the} non-uniformity of NDP, \camtwo{and thus} would cause excessive traffic across NDP units of the system upon contention (Section~\ref{Flat}).}

Overall, NDP architectures have several important characteristics that necessitate a new approach to support efficient synchronization. First, most NDP architectures~\cite{Nair2015Active,Ahn2015Scalable,Hsieh2016accelerating,Pugsley2014NDC,Youwei2019GraphQ,Boroumand2018Google,Gao2016HRL,Drumond2017mondrian,Liu2018Processing,Gao2015Practical,Gao2017Tetris,choe2019concurrent,Gokhale2015Near,Zhang2018GraphP,fernandez2020natsa,Mutlu2020AMP,Mutlu2019Processing,Ghose2019Workload,Gu2020IPIM} lack shared caches that can enable low-cost communication and synchronization among NDP cores of the system. Second, \camone{hardware cache coherence protocols are typically \camtwo{not} supported in NDP \camtwo{systems}}~\cite{Ahn2015Scalable,Hsieh2016accelerating,Pugsley2014NDC,Youwei2019GraphQ,Boroumand2018Google,Gokhale2015Near,Gao2016HRL,Drumond2017mondrian,Liu2018Processing,Gao2015Practical,Gao2017Tetris,choe2019concurrent,Kim2016Neurocube,Zhang2018GraphP,fernandez2020natsa,Mutlu2019Processing,Gu2020IPIM}, due to high area and traffic
overheads \camtwo{associated with such protocols}~\cite{Tsai2018Adaptive,Ghose2019Workload}. Third, NDP systems are non-uniform, distributed architectures, in which inter-unit communication is more expensive (both in performance and energy) than intra-unit communication~\cite{Zhang2018GraphP,Boroumand2019Conda,boroumand2017lazypim,Drumond2017mondrian,Youwei2019GraphQ,Ahn2015Scalable,Gao2015Practical,Kim2013memory}.

In this work, we present \myName{}, 
an efficient synchronization mechanism for NDP architectures. \myName{} is designed to achieve the goals of performance, cost, programming ease, and generality to cover a wide range of synchronization primitives through four key techniques. First, we offload synchronization among NDP cores to dedicated low-cost hardware units, called \myEngine{}s (\myEngineShort{}s). This approach avoids the need for complex coherence protocols and expensive \emph{rmw} operations, \camone{at} low hardware cost. Second, we directly buffer the synchronization variables in a specialized cache memory structure to avoid \camtwo{costly} memory accesses for synchronization. Third, \myName{} coordinates synchronization with a hierarchical \mpsync{} scheme: NDP cores only communicate with their local \myEngineShort{} that is located \camone{in} the same NDP unit. At the next level of communication, all \camtwo{local} \myEngineShort{}s of the system's NDP units communicate with each other to coordinate synchronization at a global level. \camtwo{Via} \camthree{its} hierarchical communication protocol, \myName{} significantly reduces synchronization traffic across NDP units \camtwo{under} \camone{high-contention} scenarios. Fourth, when applications with frequent synchronization oversubscribe the hardware synchronization resources, \myName{} uses an efficient and programmer-transparent overflow management scheme that avoids costly fallback solutions and minimizes overheads.

We evaluate \myName{} using a wide range of parallel workloads including pointer chasing, graph applications, and time series analysis. Over prior approaches (similar to~\cite{Ahn2015Scalable,Gao2015Practical}), \myName{} improves performance by 1.27$\times$ on average \camone{(up to 1.78$\times$)} under \camone{high-contention} scenarios, and by \camone{1.35$\times$ on average (up to 2.29$\times$) \camtwo{under} low-contention scenarios}. In real applications with \camtwo{fine-grained} synchronization, \myName{} comes within 9.5\% of the performance and \camtwo{6.2\%} of the energy of an ideal \camone{zero-overhead} synchronization mechanism. \camtwo{Our proposed hardware unit incurs very modest area and \camfour{power} \camthree{overheads} (Section~\ref{Areabl}) \camthree{when integrated into the compute die of an NDP unit}.}

\vspace{-1pt}
This paper makes the following contributions:
\squeezeme
\begin{itemize}[noitemsep,topsep=0pt,leftmargin=8pt]
    \item We investigate the challenges of providing efficient synchronization in Near-Data-Processing architectures, and propose an end-to-end mechanism, \myName{}, for such systems.
    \item We design low-cost synchronization units that coordinate synchronization \camone{across} NDP cores, and directly buffer synchronization variables \camone{to avoid} \camtwo{costly} memory accesses \camone{to them}. We propose an efficient \mpsync{} synchronization approach that organizes the process hierarchically, and \camone{provide a hardware-only \camtwo{programmer-transparent} overflow management scheme to alleviate performance overheads when hardware synchronization resources are exceeded.}
    \item We evaluate \myName{} using a wide range of parallel workloads and demonstrate that it significantly outperforms prior approaches both in performance and energy consumption. \camthree{\myName{} also has low hardware area \camfour{and power \camnine{overheads}.}}
\end{itemize}
 
\section{Background and Motivation}
\label{Motivationbl}

\subsection{Baseline Architecture}

\begin{figure}[H]
  \centering
   \includegraphics[width=0.9\linewidth]{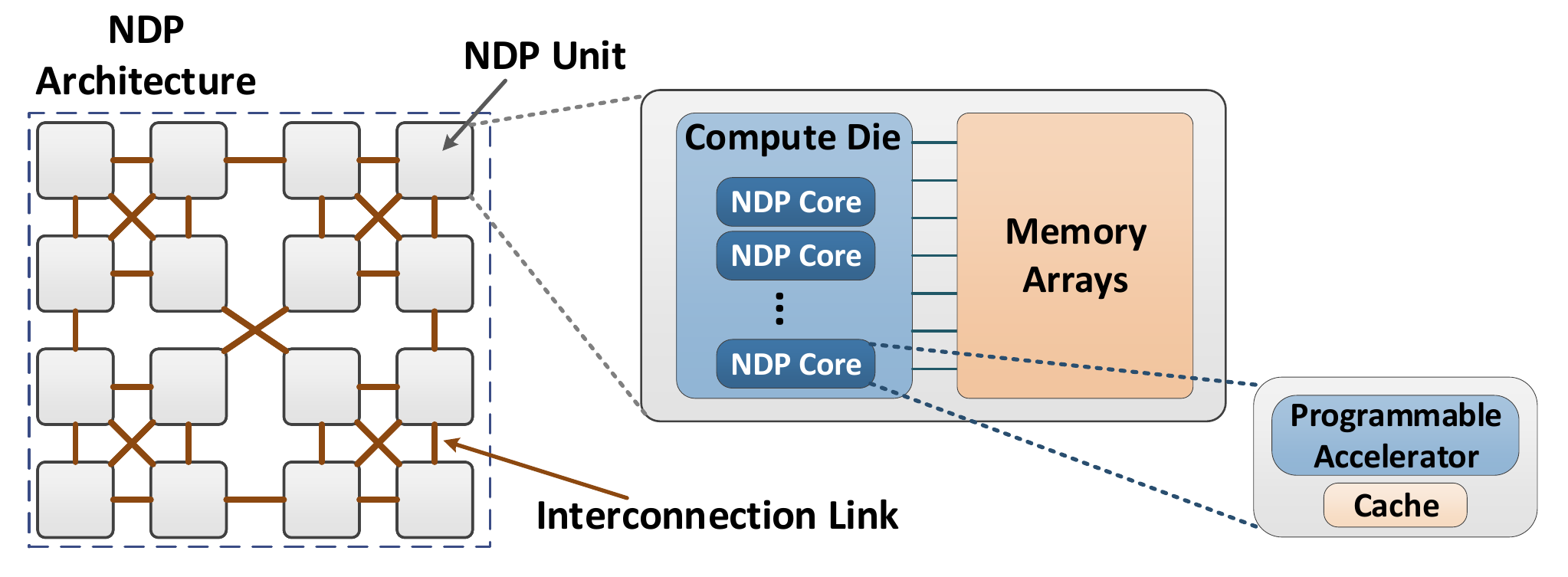}
  \caption{High-level organization of an NDP architecture.}
  \label{fig:PIMArch}
  \vspace{-2pt}
\end{figure}

Numerous works~\cite{Hsieh2016accelerating,Ahn2015Scalable,Gao2015Practical,Nai2017GraphPIM,Youwei2019GraphQ,Zhang2018GraphP,Gao2017Tetris,Kim2016Neurocube,Gu2016Leveraging,Drumond2017mondrian,Boroumand2018Google,Ahn2015PIMenabled,Boroumand2019Conda,boroumand2017lazypim,Gu2020IPIM,Tsai2018Adaptive,liu2017concurrent,choe2019concurrent,Seshadri2017Ambit,Kanellopoulos2019SMASH} show the potential benefit of NDP \camone{for} parallel, irregular applications. These proposals focus on the design of the compute logic that is placed close to or within memory, and in many cases \camtwo{provide} special-purpose near-data accelerators for specific applications. Figure~\ref{fig:PIMArch} shows the baseline organization of the NDP architecture we assume in this work, which includes several NDP units connected with each other \camone{via} serial interconnection links to share the same physical address space. Each NDP unit includes the memory arrays and a compute die with multiple low-power programmable cores or fixed-function accelerators, which we henceforth refer to as NDP cores. NDP cores execute the offloaded NDP kernel and access the various memory locations across NDP units with non-uniform access times~\cite{Zhang2018GraphP,Tsai2018Adaptive,Ahn2015Scalable,Youwei2019GraphQ,Boroumand2019Conda,boroumand2017lazypim,Drumond2017mondrian}. \camone{We assume that there is no OS running in the NDP system.} In our evaluation, we use programmable in-order NDP cores, each including small private L1 I/D caches. However, \myName{} can be used with any programmable, fixed-function or reconfigurable NDP accelerator. We assume software-assisted cache-coherence (provided by the operating system or the programmer), similar to~\cite{Tsai2018Adaptive,Gao2015Practical}: data can be either thread-private, shared read-only, or shared read-write. Thread-private and shared read-only data can be cached by NDP cores, while shared read-write data is uncacheable.

We focus on three characteristics of NDP architectures that are of particular importance in the synchronization context. First, NDP architectures typically do not have a shared level of cache memory~\cite{Nair2015Active,Ahn2015Scalable,Hsieh2016accelerating,Pugsley2014NDC,Youwei2019GraphQ,Boroumand2018Google,Gao2016HRL,Drumond2017mondrian,Liu2018Processing,Gao2015Practical,Gao2017Tetris,choe2019concurrent,Gokhale2015Near,Zhang2018GraphP,fernandez2020natsa,Mutlu2020AMP,Mutlu2019Processing,Ghose2019Workload,Gu2020IPIM}, since the NDP-suited workloads \camtwo{usually} do not benefit from deep cache hierarchies due to their poor locality~\cite{Gao2015Practical,Gagandeep2019Near,Tsai2018Adaptive,oliveira2021pimbench}. Second, NDP architectures do not typically support conventional hardware cache coherence protocols~\cite{Ahn2015Scalable,Hsieh2016accelerating,Pugsley2014NDC,Youwei2019GraphQ,Boroumand2018Google,Gokhale2015Near,Gao2016HRL,Drumond2017mondrian,Liu2018Processing,Gao2015Practical,Gao2017Tetris,choe2019concurrent,Kim2016Neurocube,Zhang2018GraphP,fernandez2020natsa,Mutlu2019Processing,Gu2020IPIM}, because they would add area and traffic \camfive{overheads~\cite{Tsai2018Adaptive,Ghose2019Workload}, and} would incur high complexity and latency~\cite{Abeydeera2020Chronos}, limiting the benefits of NDP. Third, communication across NDP units is expensive, because NDP systems are non-uniform distributed architectures. The energy and performance costs of inter-unit communication are typically orders of magnitude greater than the costs of intra-unit communication~\cite{Zhang2018GraphP,Boroumand2019Conda,boroumand2017lazypim,Drumond2017mondrian,Youwei2019GraphQ,Ahn2015Scalable,Gao2015Practical,Kim2013memory}, and \camone{thus inter-unit communication may \camfive{slow down}} the execution of NDP cores~\cite{Zhang2018GraphP}.


\subsection{The Solution Space for Synchronization}
Approaches to support synchronization are typically either \camone{via} shared memory or \mpsync{} schemes.

\subsubsection{Synchronization \camone{via} \camone{S}hared \camone{M}emory} In this case, cores coordinate 
via a consistent view of shared memory locations, 
using atomic read/write operations or atomic read-modify-write (\emph{rmw}) operations. If \emph{rmw} operations are \camfour{\emph{not}} supported by hardware, \camone{Lamport's bakery} algorithm~\cite{Lamport1974New} can provide synchronization to $N$ participating cores, assuming sequential consistency~\cite{Lamport1979How}. However, this scheme scales poorly, as a core accesses $O(N)$ memory locations at \emph{each} synchronization retry.
\camfive{In contrast, commodity systems (CPUs, GPUs, MPPs) typically support \emph{rmw} operations in hardware.}

\camfour{GPUs and MPPs support \emph{rmw} operations} \camone{in} specialized hardware units (known as \emph{remote atomics}), located \camone{in} each bank of the shared cache~\cite{Wittenbrink2011Fermi,Luna2013Performance}, or the memory controllers~\cite{kessler1993crayTA,Laudon1997SGI}.
\camtwo{Remote atomics are also supported by an NDP work~\cite{Gao2015Practical} at the vault controllers of Hybrid Memory Cube (HMC)~\cite{HMC,HMC_old}.}
Implementing synchronization primitives \camone{using} remote atomics requires a spin-wait scheme, i.e., executing consecutive \emph{rmw} retries. However, performing and sending every \emph{rmw} operation to a shared, fixed location \camone{can} cause high global traffic and create hotspots~\cite{Wang2019Fast,Mukkara2019PHI,li2015fine,yilmazer2013hql,eltantawy2018warp}. \camone{In NDP systems}, consecutive \emph{rmw} operations to a remote NDP unit would incur high traffic \emph{across} NDP units, with high performance and energy overheads.

Commodity CPU architectures support \emph{rmw} operations either by locking the bus (or equivalent link), or by relying on the hardware cache coherence protocol~\cite{Sorin2011Primer,intelsys}, which \camone{many} NDP architectures do not support. \camtwo{Therefore}, coherence-based synchronization~\cite{guiroux2016multicore,rudolph1984dynamic,anderson1989performance,mellor1991algorithms,scott2002non,Dice2015Lock,magnusson1994queue,craig1993building,luchangco2006hierarchical,dice2011flat,chabbi2015high,Zhang2016Scalable} cannot be directly implemented in NDP architectures. Moreover, based on prior works on synchronization~\cite{David2013Everything,Boyd2010AnAnalysis,Kaxiras2015Turning,Mellor1991Synchronization,Tallent2010Analyzing,Molka2009Memory}, coherence-based synchronization would exhibit low scalability on NDP systems for two reasons. First, it performs poorly with a \emph{\camtwo{large}} number of cores, due to low scalability of conventional hardware coherence protocols~\cite{Heinrich1999Aquantitative,Sorin2011Primer,Kelm2010Cohesion,Kelm2010Waypoint}. Most NDP systems include several NDP units~\cite{Kim2013memory,Zhang2018GraphP,Youwei2019GraphQ,Ahn2015Scalable}, each typically supporting hundreds of \camtwo{small, area-constrained} cores~\cite{Ahn2015Scalable, Boroumand2018Google,Youwei2019GraphQ,Zhang2018GraphP}. Second, the non-uniformity in memory accesses significantly affects the scalability of coherence-based synchronization~\cite{David2013Everything,Boyd2010AnAnalysis,Zhang2016Scalable,Molka2009Memory}. Prior work on coherence-based synchronization~\cite{David2013Everything} observes that the latency of a lock acquisition that needs to transfer the lock \emph{across} \camtwo{NUMA} sockets can be up to 12.5$\times$ \camtwo{higher} than that \emph{within} a socket. We expect such effects to be aggravated in NDP systems, since they are by nature \emph{non-uniform} and \emph{distributed}~\cite{Kim2013memory,Zhang2018GraphP,Youwei2019GraphQ,Ahn2015Scalable,Gao2015Practical,Boroumand2019Conda,boroumand2017lazypim,Drumond2017mondrian} \camone{with very low memory access latency within an NDP unit.}

We validate these observations on both a real CPU and our simulated NDP system. \camfour{On an Intel Xeon Gold server, we evaluate the} \camtwo{operation} throughput achieved by two coherence-based lock algorithms (Table~\ref{Tab:motivation}), i.e., TTAS~\cite{rudolph1984dynamic} and Hierarchical Ticket Lock (HTL)~\cite{mellor1991algorithms}, using a microbenchmark taken from \camone{the} \emph{libslock} library~\cite{David2013Everything}. \camone{When increasing the number of threads from 1 to 14 within a single socket, throughput drops by 3.91$\times$ and 2.77$\times$ for TTAS and HTL, respectively. Moreover, when pinning \camtwo{two} threads on different NUMA sockets, \camtwo{throughput} drops by up to 2.29$\times$ over when pinning them on the same socket, due to non-uniform memory access times \camtwo{of} lock variables.}

\vspace{3pt}
\begin{table}[H]
\centering
\resizebox{0.99\columnwidth}{!}{%
\begin{tabular}{c||c c||c c}
\toprule
Million Operations & 1 thread & 14 threads & 2 threads & 2 threads\\
per Second & single-socket & single-socket & same-socket & different-socket \\
\midrule
TTAS lock~\cite{rudolph1984dynamic}                & 8.92  & 2.28 & 9.91 & 4.32                     \\
Hierarchical Ticket lock~\cite{mellor1991algorithms} & 8.06 & 2.91 & 9.01 & 6.79 \\
\bottomrule

\end{tabular}%
}
\caption{Throughput of two coherence-based lock algorithms on an Intel Xeon Gold \camtwo{server} using \camone{the} libslock library~\cite{David2013Everything}.}
\label{Tab:motivation}
\vspace{-2pt}
\end{table}

\begin{figure}[H]
    \centering\captionsetup[subfloat]{labelfont=bf}
  \begin{subfigure}[h]{0.46\columnwidth}
    \hspace{-14pt}
    \includegraphics[scale=0.154]{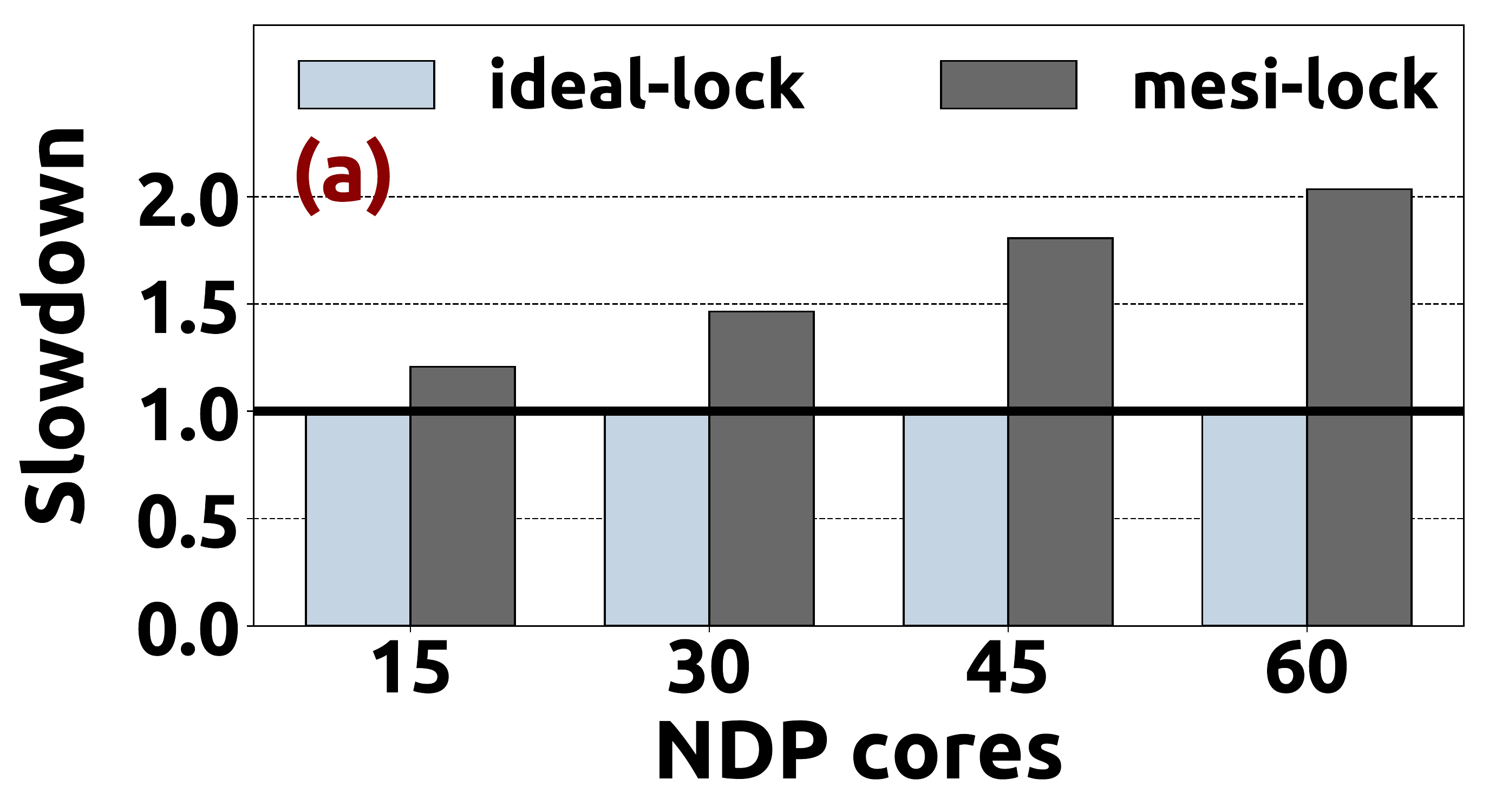}
    \vspace{-12pt}
    \label{fig:coherence} 
 \end{subfigure}
  ~
  \begin{subfigure}[h]{0.46\columnwidth}
   \centering
    \includegraphics[scale=0.154]{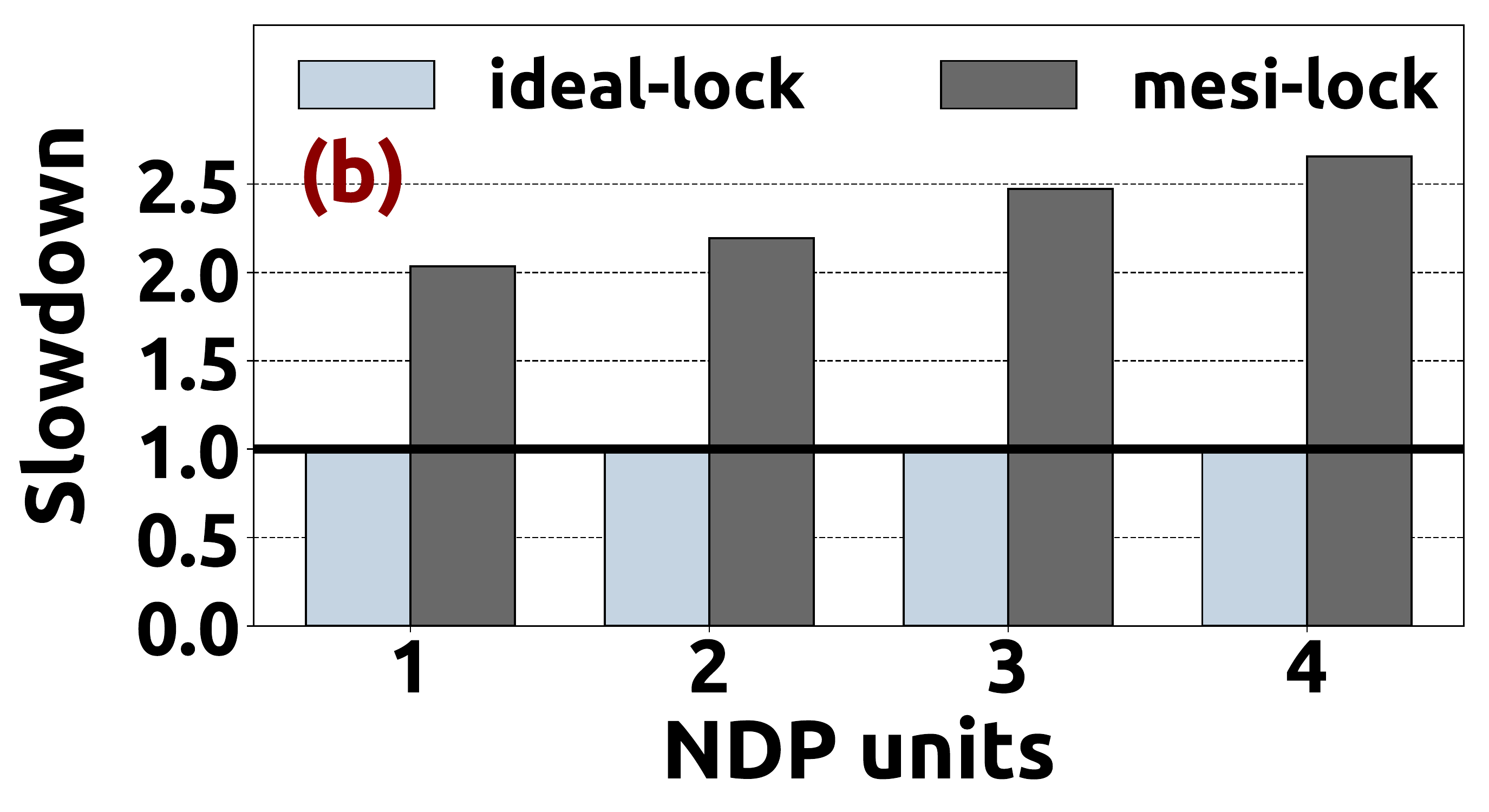}
    \vspace{-12pt}
    \label{fig:numaness}
  \end{subfigure} 
  \caption{\camone{\camtwo{Slowdown} of a stack data structure using a coherence-based lock over using an \emph{ideal} zero-cost lock,} when varying (a) the NDP cores \camfive{within} a single NDP unit and (b) the number of NDP units \camthree{while keeping core count constant at 60}.}
  \label{fig:motcoh}
  \vspace{-4pt}
\end{figure}

In \camfour{our simulated NDP system,} we evaluate the \camtwo{performance} achieved by a stack data structure protected with a \camtwo{coarse-grained} lock. Figure~\ref{fig:motcoh} shows the \camtwo{slowdown} of the stack when using a coherence-based lock~\cite{herlihy2008art} (\emph{mesi-lock}), implemented upon a MESI directory coherence protocol, over using an ideal lock with zero cost for synchronization (\emph{ideal-lock}). \camfour{First, we observe that the \camtracy{high contention for} the cache line containing the \emph{mesi-lock} and the resulting coherence traffic inside the network significantly limit scalability of the stack as the number of cores increases. With 60 NDP cores within a single NDP unit (Figure~\ref{fig:motcoh}a), the stack with \emph{mesi-lock} incurs 2.03$\times$ slowdown over \emph{ideal-lock}. Second, we notice that the non-uniform memory accesses to the cache line containing the \emph{mesi-lock} also impact the scalability of the stack. When increasing the number of NDP units while keeping total core count constant at 60 (Figure~\ref{fig:motcoh}b), the slowdown of the stack with \emph{mesi-lock} increases to 2.66$\times$ (using 4 NDP units) over \emph{ideal-lock}.} In \emph{non-uniform} NDP systems, the scalability of coherence-based synchronization is severely limited by the long transfer latency and low bandwidth of the interconnect used between the NDP units.

\subsubsection{\Mpsync{} \camone{S}ynchronization}
In this \camone{approach}, cores coordinate \camone{with each other} by exchanging messages (either in software or hardware) \camone{in order to} reach an agreement. For instance, a recent NDP work~\cite{Ahn2015Scalable} implements a barrier primitive via hardware message-passing communication among NDP cores, \camone{i.e., one core of the system works as a \emph{master} core to collect the synchronization status of the \camtwo{rest}.}
To \camone{improve system performance in \emph{non-uniform} HMC-based NDP systems}, Gao et al.~\cite{Gao2015Practical} propose a \emph{tree-style} barrier primitive, \camone{where cores exchange messages to first synchronize within \camtwo{a} vault, then across the vaults of \camtwo{an} HMC cube, and finally across HMC cubes.} In general, optimized \mpsync{} synchronization \camone{schemes proposed} in the literature~\cite{abellan2010g, Tang2019plock,Hoefler2004survey, hensgen1988two,grunwald1994efficient,Gao2015Practical} aim to minimize (i) the number of messages sent among cores, and (ii) expensive network traffic. \camtwo{To avoid} the \camone{major} issues of synchronization \camone{via} shared memory described above, we design our approach building \camtwo{on} the \mpsync{} synchronization concept.


\section{\myName{}: Overview}
\label{Overviewbl}

\myName{} is an end-to-end solution for synchronization in NDP architectures that improves performance, has low cost, eases programmability, and supports multiple synchronization primitives.
\myName{} relies on the following key techniques:

\noindent \textbf{1. Hardware support for synchronization acceleration:}
We design low-cost hardware units, \camone{called} \myEngine{}s (\myEngineShort{}s), to coordinate the synchronization among NDP cores of the system. \myEngineShort{}s eliminate the need for complex cache coherence protocols and expensive \emph{rmw} operations, and incur modest hardware cost.

\noindent \textbf{2. Direct buffering of synchronization variables:}
We add a specialized cache structure, the Synchronization Table (\myTableShort{}), inside \camone{an} \myEngineShort{} to keep synchronization information. \camtwo{Such direct} buffering avoids \camtwo{costly} memory accesses for synchronization, and enables high performance under \camone{low-contention} scenarios.

\noindent \textbf{3. Hierarchical message-passing communication:}
We organize the communication hierarchically, 
with each NDP unit including an \myEngineShort{}. NDP cores communicate with their local \myEngineShort{} that is located \camone{in} the same NDP unit. \myEngineShort{}s communicate with each other to coordinate synchronization at a global level. Hierarchical communication minimizes expensive communication \emph{across} NDP units, and achieves high performance under \camone{high-contention} scenarios.

\noindent \textbf{4. Integrated hardware-only overflow management:} 
We incorporate a hardware-only overflow management scheme to efficiently handle scenarios when \myTableShort{} is fully occupied. This \camtwo{programmer-transparent} technique \camtwo{effectively limits} performance degradation \camtwo{under} overflow scenarios.

\subsection{Overview of \myName{}}

\begin{figure}[H]
  \hspace{-4pt}\includegraphics[scale=0.4]{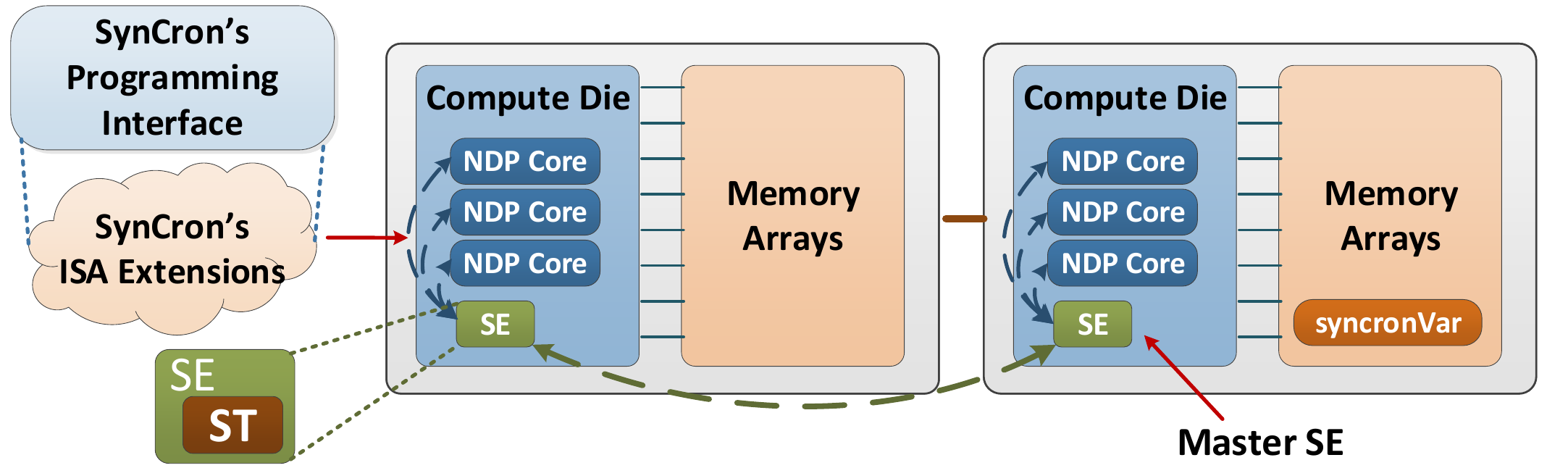}
  \caption{High-level overview of \myName{}.}
  \label{fig:overview}
\end{figure}

Figure~\ref{fig:overview} provides an overview of our approach. \myName{} exposes a simple programming interface such that programmers can easily \camone{use a variety of synchronization primitives in} their multithreaded applications \camone{when \camtwo{writing} them \camtwo{for}} NDP \camone{systems}. The interface is implemented using \camtwo{two new instructions} that are used by NDP cores to communicate synchronization requests to \myEngineShort{}s. These are general \camone{enough} to cover all semantics for the most widely-used synchronization primitives.

We add one \myEngineShort{} \camone{in} the compute die of each NDP unit. For a particular synchronization variable allocated \camone{in} an NDP unit, the \myEngineShort{} that is physically located \camone{in} the same NDP unit is considered the \masterSE{}. In other words, the \masterSE{} is defined by the address of the synchronization variable. \camtwo{It} 
is responsible for the global coordination of synchronization on that variable, i.e., among all \myEngineShort{}s of the system. All other \myEngineShort{}s are responsible \camtwo{only} for the local coordination of synchronization among the cores \camone{in} the same NDP unit with them. 

NDP cores act as clients that send requests to \myEngineShort{}s via hardware message-passing. \myEngineShort{}s act as servers that process synchronization requests. In the proposed hierarchical communication, NDP cores send requests to their local \myEngineShort{}s, while \myEngineShort{}s of different NDP units communicate with the \masterSE{} of the specific variable, to coordinate the process at a global level, i.e., among all NDP \camone{units}. 

When an \myEngineShort{} receives a request from an NDP core for a synchronization variable, it directly buffers \camthree{the} variable \camone{in its} \myTableShort{}, keeping all the information needed for synchronization \camone{in the \myTableShort{}}. \camtwo{If the \myTableShort{} is full, we use the main memory as a fallback solution. To hierarchically coordinate synchronization via main memory in \myTableShort{} overflow cases, we design (i) a generic structure, called \emph{\mySyncVar{}}, to keep track of required synchronization information, and (ii) specialized \emph{overflow} messages to be sent among \myEngineShort{}s. 
The hierarchical communication among \myEngineShort{}s is implemented via corresponding support in message encoding, the \myTableShort{}, and \emph{\mySyncVar{}} \camthree{structure}.}

\subsection{\myName{}'s Operation} 

\myName{} supports locks, barriers, semaphores, and condition variables.
Here, we present \myName{}'s operation for locks. \myName{} has similar behavior for the \camone{other} \camtwo{three} primitives.

\begin{figure}[H]
  \centering
  \includegraphics[scale=0.45]{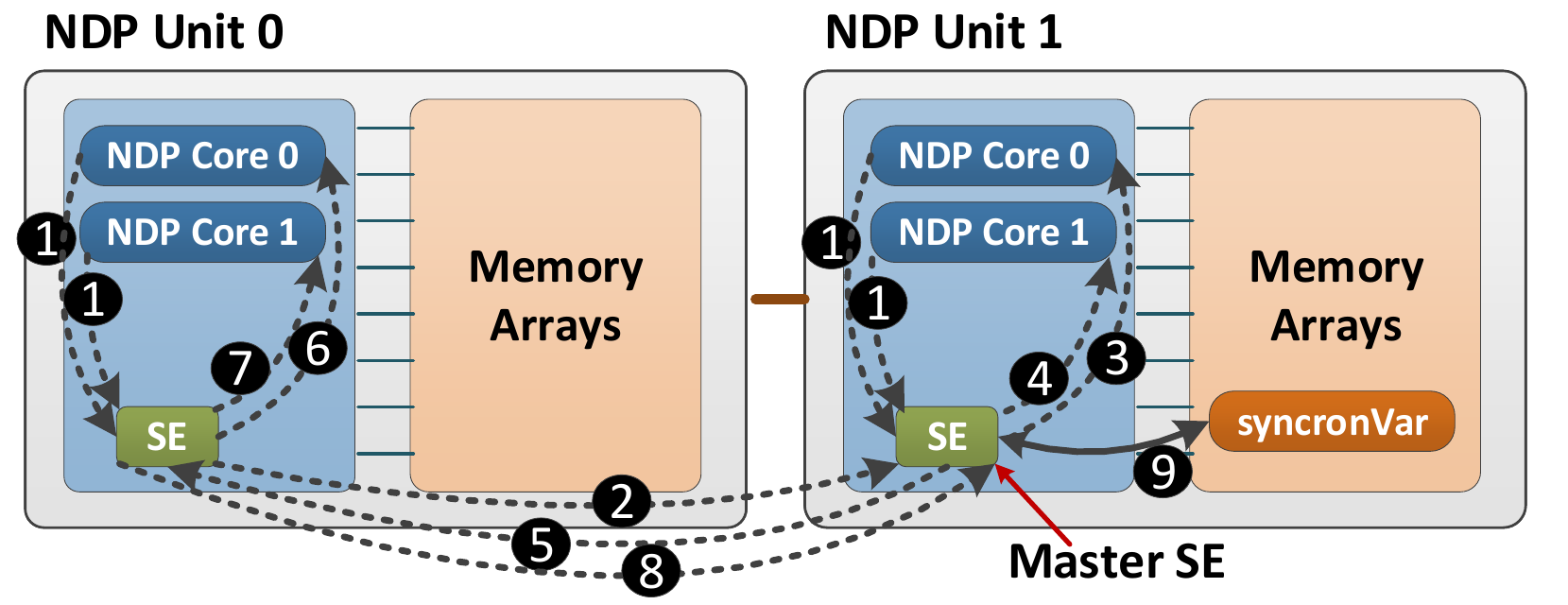}
  \caption{An example execution scenario for a lock \camone{requested by \emph{all} \camtwo{NDP} cores}.}
  \label{fig:lock}
\end{figure}

\noindent
\textbf{Lock Synchronization Primitive:}
Figure~\ref{fig:lock} \camone{shows} a system 
\camone{composed} \camtwo{of} two NDP units \camone{with} two NDP cores each. \camone{In this example,} all cores \camone{request and} compete for the same lock. \camone{First,} all NDP cores send \emph{local} lock acquire messages to their \myEngineShort{}s \circled{1}. \camone{After receiving these messages,} \camone{each} \myEngineShort{} \camone{keeps track of its requesting cores by reserving} one new entry in \camone{its} \myTableShort{}, i.e., \camone{directly buffering the lock variable in \myTableShort{}.} Each \myTableShort{} entry includes \camone{a local waiting list (i.e.,} a hardware bit queue with one bit for each local NDP core), and \camone{a global waiting list (i.e.,} a bit queue with one bit for each \myEngineShort{} of the system). 
\camone{To keep track of the requesting cores,} \camone{each} \myEngineShort{} sets \camone{the bits} corresponding to the \camone{requesting} cores in the local waiting list of the \myTableShort{} entry. When the local \myEngineShort{} receives a request for a synchronization variable \emph{for the first time}, it sends a \emph{global} lock acquire message to the \masterSE{} \circled{2}, which in turn sets the corresponding bit in the global waiting list \camone{in its \myTableShort{}}. 
\camone{This way, the \masterSE{} keeps track of all requests to a particular variable coming from an \myEngineShort{}, and can arbitrate \camtwo{between} different \myEngineShort{}s.} 
\camone{The local \myEngineShort{} can then serve successive local requests to the same variable until there are no other local requests.}
\camone{By using} the \camone{proposed} hierarchical communication protocol, the cores send local messages to their local \myEngineShort{}, and the \myEngineShort{} \camone{needs to send} \emph{only one aggregated} message, on behalf of all its local waiting cores, \camone{to} the \masterSE{}. 
\camone{As a result, we reduce the need for communication} through the narrow, expensive links that connect \camone{different} NDP units.

The \masterSE{} first prioritizes the local waiting list, granting the lock to \camone{its own} local NDP cores \camone{in sequence} \camone{(e.g., to NDP \camthree{Core} 0 \camtwo{first} \circled{3}, \camtwo{and} to NDP \camthree{Core} 1 \camtwo{next} \circled{4} in Figure~\ref{fig:lock})}. \camtwo{At the end of the critical section,} \camone{each} local lock owner sends a lock release message to \camone{its} \myEngineShort{} \camone{in order to release the lock.} 
When there are no other local requests, the \masterSE{} transfers the control of the lock to the \myEngineShort{} of another NDP unit based on \camone{its} global waiting list \circled{5}. Then, the local \myEngineShort{} grants the lock \camone{to} \camone{its} local NDP cores \camone{in sequence} \camfour{(e.g., \circled{6}, \circled{7})}. After all local cores \camone{release the lock}, \camone{the \myEngineShort{}} sends an \emph{aggregated} global lock release message to \camone{the} \masterSE{} \circled{8} 
and releases its \myTableShort{} entry. When the message arrives \camone{at} the \masterSE{}, if there \camone{are no other pending requests to the same variable}, \camone{the \masterSE{}} releases its \myTableShort{} entry. \camone{In this example, \myEngineShort{}s directly buffer the lock variable in their \myTableShort{}s. If an \myTableShort{} is \emph{full}, the \masterSE{} globally coordinates synchronization by keeping track \camtwo{of} all \camtwo{required} information in main memory \circled{9}, via \camtwo{our} proposed \camfour{overflow management scheme (Section~\ref{Overflowbl}).}
}

\section{\myName{}: Detailed Design}
\label{Mechanismbl}

\myName{} leverages the key observation that all synchronization primitives fundamentally communicate the same information, i.e., a waiting list of cores that participate in the synchronization 
process, and a condition to be met to notify one or more cores. Based on this observation, we design \myName{} to cover the four most widely used synchronization primitives.
Without loss of generality, we assume that each NDP core represents a hardware thread \camtwo{context} with a unique ID. To support multiple hardware thread \camtwo{contexts} per NDP core, the corresponding hardware structures of \myName{} \camone{need} to be augmented to include 1-bit per hardware thread \camtwo{context}.

\subsection{Programming Interface and ISA \camtwo{E}xtensions}\label{SynCronISAbl}

\myName{} provides lock, barrier, semaphore and condition variable synchronization primitives, \camfour{supporting} two types of barriers: within cores of the \emph{same} NDP unit and within cores across \camtwo{different} NDP units of the system. 
\camtwo{\myName{}'s} programming interface (\camone{Table~\ref{tab:interface}}) implements the synchronization semantics with two new ISA instructions, \camone{which} are \emph{rich} and \emph{general} \camone{enough} to express all supported primitives. NDP cores use these instructions to assemble messages for synchronization requests, which are issued through the network to \myEngineShort{}s.

\begin{table}[H]
\centering
  \begin{minipage}{.4\textwidth}
  \centering
  \resizebox{\textwidth}{!}{
    \begin{tabular}{l} 
    \toprule
    \textbf{\camtwo{\myName{} Programming Interface}} \\ 
    \midrule
    \textcolor{blue(pigment)}{\camone{\mySyncVar}} *create\_syncvar (); \\
    \textcolor{blue(pigment)}{void} destroy\_syncvar (\textcolor{blue(pigment)}{\camone{\mySyncVar}} *svar); \\
    \textcolor{blue(pigment)}{void} lock\_acquire (\textcolor{blue(pigment)}{\camone{\mySyncVar}} *lock); \\
    \textcolor{blue(pigment)}{void} lock\_release (\textcolor{blue(pigment)}{\camone{\mySyncVar}} *lock); \\
    \textcolor{blue(pigment)}{void} barrier\_wait\_within\_unit (\textcolor{blue(pigment)}{\camone{\mySyncVar}} *bar, \textcolor{blue(pigment)}{int} initialCores); \\
    \textcolor{blue(pigment)}{void} barrier\_wait\_across\_units (\textcolor{blue(pigment)}{\camone{\mySyncVar}} *bar, \textcolor{blue(pigment)}{int} initialCores); \\
    \textcolor{blue(pigment)}{void} sem\_wait (\textcolor{blue(pigment)}{\camone{\mySyncVar}} *sem, \textcolor{blue(pigment)}{int} initialResources); \\
    \textcolor{blue(pigment)}{void} sem\_post (\textcolor{blue(pigment)}{\camone{\mySyncVar}} *sem);\\
    \textcolor{blue(pigment)}{void} cond\_wait (\textcolor{blue(pigment)}{\camone{\mySyncVar}} *cond, \textcolor{blue(pigment)}{\camone{\mySyncVar}} *lock); \\
    \textcolor{blue(pigment)}{void} cond\_signal (\textcolor{blue(pigment)}{\camone{\mySyncVar}} *cond);\\ 
    \textcolor{blue(pigment)}{void} cond\_broadcast (\textcolor{blue(pigment)}{\camone{\mySyncVar}} *cond);\\
    \bottomrule
    \end{tabular}}
  \end{minipage}
   \caption{\label{tab:interface}\myName{}'s Programming Interface \camtwo{(i.e., API)}.}
   \vspace{-8pt}
\end{table}

\textbf{\textit{req\_sync \camtwo{addr, opcode, info}}}: This instruction creates a message and commits when a response message is received back. The \camtwo{\emph{addr}} register has the address of a synchronization variable, the \camtwo{\emph{opcode}} register has the message \camtwo{opcode of a particular semantic of \camthree{a} synchronization primitive (Table~\ref{tab:opcodes})}, and the \camtwo{\emph{info}} register has specific information needed for \camthree{the} primitive (\emph{MessageInfo} in message encoding of Fig.~\ref{fig:msgencoding}).

\textbf{\textit{req\_async \camtwo{addr, opcode}}}: This instruction creates a message and after the message is issued \camthree{to} the network, the instruction commits. The registers \camtwo{\emph{addr}, \emph{opcode}} have the same \camtwo{semantics} as in \emph{req\_sync} instruction.

\subsubsection{Memory Consistency} We design \myName{} assuming a relaxed consistency memory model. The proposed ISA extensions \camtwo{act as memory fences.}
First, \emph{req\_sync}, commits once a message (ACK) is received \camtwo{(from the local \myEngineShort{} to the core)}, which ensures that all following instructions will be issued after \emph{req\_sync} has been completed. Its semantics is similar \camtwo{to those of} the SYNC and ACQUIRE operations of Weak Ordering (WO)~\cite{Culler1999Parallel} and Release Consistency (RC)~\cite{Culler1999Parallel} models, respectively. Second, \emph{req\_async}, does not require a return message (ACK). It is issued once all previous instructions \camtwo{are} completed. \camtwo{Its} semantics is similar \camtwo{to \camthree{that} of} the RELEASE operation of RC~\cite{Culler1999Parallel}. \camtracy{In the case} of WO, \emph{req\_sync} is sufficient. \camtracy{In the case} of RC, the \emph{req\_sync} instruction is used for \camone{acquire-type semantics, i.e., } lock\_acquire, barrier\_wait, semaphore\_wait and condition\_variable\_wait, while the \emph{req\_async} instruction is used for \camone{release-type semantics, i.e.,} lock\_release, semaphore\_post, condition\_variable\_signal, and condition\_variable\_broadcast.

\subsubsection{Message Encoding}\label{Encodingbl}

\begin{figure}[H]
  \vspace{2pt}
  \centering
  \includegraphics[width=0.9\columnwidth]{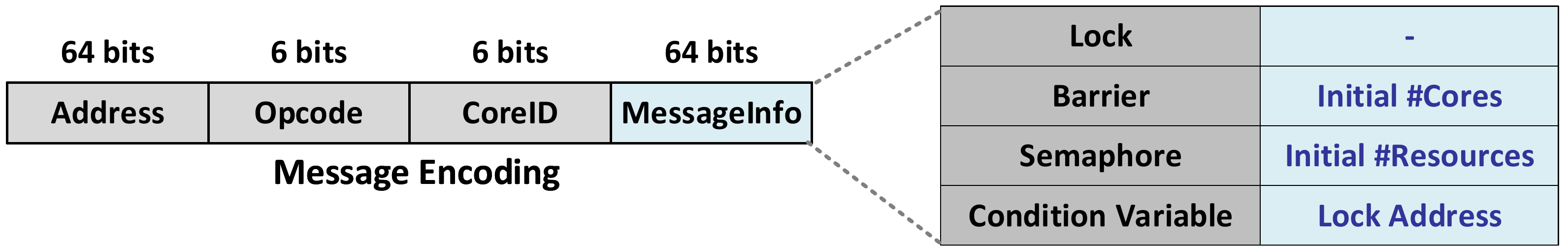}
  \caption{Message \camtwo{e}ncoding of \myName{}.} 
  \label{fig:msgencoding}
\end{figure}

Figure~\ref{fig:msgencoding} describes the encoding of the message used for communication between \camone{NDP} cores and the \camone{\myEngineShort{}}. Each message includes: (i) the 64-bit address of the synchronization variable, (ii) the message \camone{opcode} that implements the semantics of the different synchronization primitives (6 bits cover all message \camone{opcodes}), (iii) the unique \camtwo{ID} number of the NDP core (6 bits are sufficient 
\camsix{for our simulated NDP system in Section~\ref{Methodologybl}),} 
and (iv) a 64-bit field (\emph{MessageInfo}) that communicates specific information needed for each different synchronization primitive, i.e., the number of the cores that participate in a barrier, the initial value of a semaphore, the address of the lock associated with a condition variable.

\camnine{\noindent\textbf{Hierarchical Message Opcodes.}}
\myName{} enables a hierarchical scheme, where the \myEngineShort{}s \camtwo{of NDP units} communicate with each other to coordinate synchronization at a global level. Therefore, we support two types of messages (Table~\ref{tab:opcodes}): \camone{(i)} \emph{local}, which are used by NDP cores to communicate with their local \myEngineShort{}, and \camone{(ii)} \emph{global}, which are used by \myEngineShort{}s to communicate with \camone{the} \masterSE{}, and vice versa. \camtwo{Since we support two types of barriers (Table~\ref{tab:interface}), we design two message opcodes for a \emph{local} barrier\_wait message sent by an NDP
core to its local \myEngineShort{}: (i) \emph{barrier\_wait\_local\_within\_unit} is used when cores of a single NDP unit participate in the barrier, and (ii) \emph{barrier\_wait\_local\_across\_units} is used when cores from different NDP units participate in the barrier. In the latter case, if a \emph{smaller} number of cores than the total \emph{available} cores of the NDP system participate in the barrier, \myName{} supports one-level communication: local \myEngineShort{}s re-direct all messages (received from their local NDP cores) to the \masterSE{}, which globally coordinates the barrier among \emph{all} participating cores.} This design choice is a trade-off between performance (\emph{\camtwo{more} remote messages}) and hardware/ISA complexity, since \camtwo{the number of} participating cores \camtwo{of \emph{each}} NDP unit would need to be communicated to the hardware through \camtwo{additional registers in ISA,} and \camtracy{message} \camtwo{opcodes} (\emph{higher complexity}).

\begin{table}[H]
  \begin{minipage}{.50\textwidth}
  \hspace{-6pt}
  \resizebox{\textwidth}{!}{
    \begin{tabular}{c c} 
    \toprule
    \textbf{Primitives} & \textbf{\camthree{\myName{}} Message \camone{Opcodes}} \\ [0.1ex] 
    \midrule
    \midrule
    \multirow{3}{*}{\textbf{Locks}} & lock\_acquire\_global, lock\_acquire\_local, lock\_release\_global \\ 
    & lock\_release\_local,
      lock\_grant\_global,
     lock\_grant\_local \\ 
     & lock\_acquire\_overflow,  
     lock\_release\_overflow, lock\_grant\_overflow \\ 
    \hline
     \multirow{3}{*}{\textbf{Barriers}} & barrier\_wait\_global, barrier\_wait\_local\_within\_unit \\ 
     & barrier\_wait\_local\_across\_units,
    barrier\_depart\_global, barrier\_depart\_local \\ 
    & barrier\_wait\_overflow, barrier\_departure\_overflow \\
    \hline
     \multirow{3}{*}{\textbf{Semaphores}} & sem\_wait\_global, sem\_wait\_local, sem\_grant\_global \\ 
     & sem\_grant\_local,
     sem\_post\_global, sem\_post\_local \\ 
     & sem\_wait\_overflow,  
     sem\_grant\_overflow, sem\_post\_overflow \\ 
    \hline
    \multirow{4}{*}{\shortstack[l]{\textbf{Condition }\\ \textbf{Variables}}} & cond\_wait\_global, cond\_wait\_local, cond\_signal\_global \\ 
    & cond\_signal\_local,
    cond\_broad\_global, cond\_broad\_local \\ 
    & cond\_grant\_global, cond\_grant\_local,
    cond\_wait\_overflow \\ 
    & cond\_signal\_overflow,
    cond\_broad\_overflow,
    cond\_grant\_overflow  \\
    \hline
    \textbf{Other} & decrease\_indexing\_counter \\
    \bottomrule
    \end{tabular}}
  \end{minipage}
  \caption{\label{tab:opcodes}Message \camtwo{o}pcodes of \myName{}.}
\end{table}

\subsection{Synchronization Engine (\myEngineShort{})} 
\begin{figure}[H]
   \centering
  \includegraphics[scale=0.56]{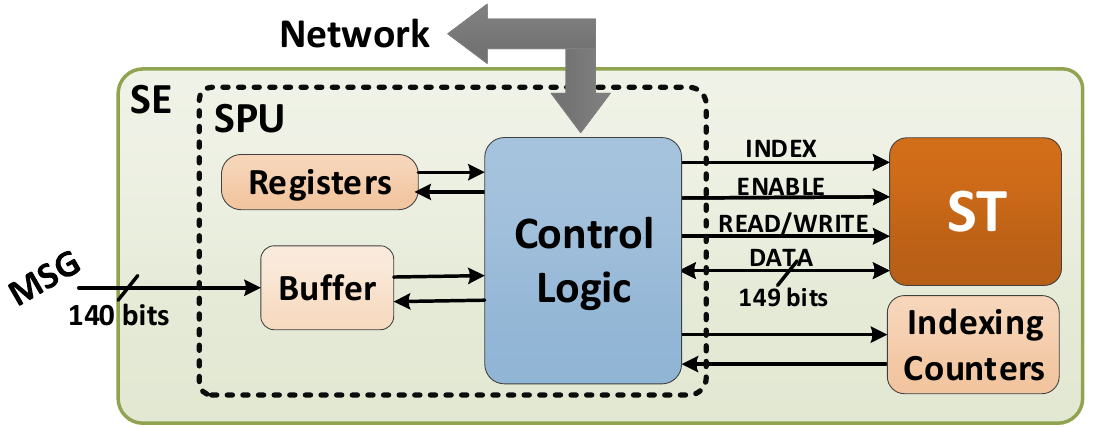}
  \caption{The Synchronization Engine \camone{(\myEngineShort{})}.}
  \label{fig:syncEng}
\end{figure}

Each \myEngineShort{} module (Figure~\ref{fig:syncEng}) is integrated into the compute die of each NDP unit. An \myEngineShort{} consists of \emph{three} components:

\subsubsection{Synchronization Processing Unit (SPU)}
The SPU is the logic that handles the messages, updates the \myTableShort{}, and issues requests to memory \camone{as} needed. The SPU includes the control unit, a buffer, and a few registers. The buffer is a small \camtwo{SRAM queue} for temporarily storing messages that arrive \camone{at the} \myEngineShort{}. The control unit implements custom logic with simple \camone{logical bitwise} operators (and, or, xor, zero) and multiplexers. 

\vspace{-4pt}
\vspace{-4pt}

\subsubsection{Synchronization Table (\myTableShort{})}
\begin{figure}[H]
  \vspace{2pt}
  \centering
  \includegraphics[width=0.98\columnwidth]{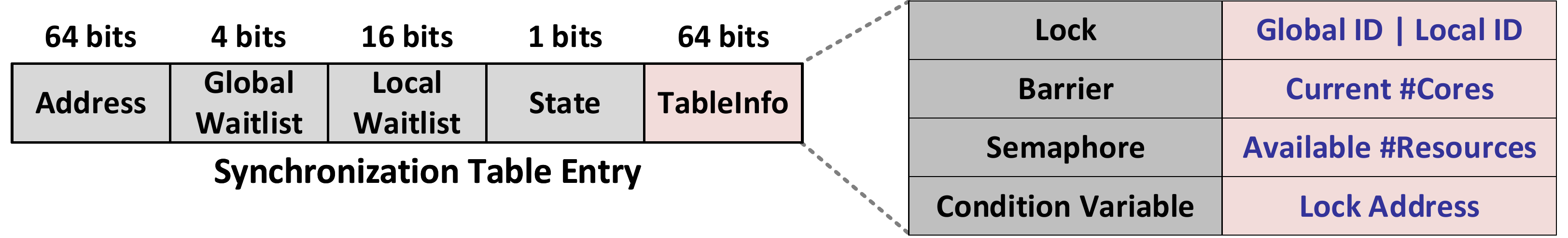}
  \caption{Synchronization Table (\myTableShort{}) \camtwo{e}ntry.}
  \label{fig:stentry}
  \vspace{-4pt}
\end{figure}

\myTableShort{} keeps track of all the information needed to coordinate synchronization. Each \myTableShort{} has 64 entries. \camone{Figure~\ref{fig:stentry} shows an} \myTableShort{} entry\camtwo{,} \camone{which} includes: (i) the 64-bit address of a synchronization variable, (ii) the global waiting list used by the \masterSE{} for global synchronization among \myEngineShort{}s, i.e., a hardware bit queue \camtwo{including one bit for} each \myEngineShort{} of the system, (iii) the local waiting list used by all \myEngineShort{}s for synchronization among the NDP cores of an NDP unit, i.e., a hardware bit queue \camtwo{including one bit for}  each NDP core within the unit, (iv) the state of the \myTableShort{} entry, \camone{which can be either \emph{free} or \emph{occupied}}, and (v) a 64-bit field (\emph{TableInfo}) to track specific information needed for each synchronization primitive. \camone{For the} lock primitive, the \emph{TableInfo} field is used to indicate \camone{the lock owner that is either an \myEngineShort{} of \camtwo{an} NDP unit (\emph{Global ID} represented by the most significant bits)  or a \camtwo{\emph{local} NDP core} (\emph{Local ID} represented by the least significant bits).} \camtwo{We} assume that all NDP cores of an NDP unit have a unique \camfive{\emph{local ID}} within the NDP unit, while all \myEngineShort{}s of the system have a unique \camfive{\emph{global ID}} within the system. \camone{The number of bits in the global and local waiting lists of Figure~\ref{fig:stentry} is specific for the configuration of our evaluated system (Section~\ref{Methodologybl}), which includes
\camthree{16 NDP cores per NDP unit and 4 \myEngineShort{}s (one per NDP unit), and has to be extended accordingly, if the system supports more NDP cores or \myEngineShort{}s.}
}


\subsubsection{Indexing Counters}
If an \myTableShort{} is full, i.e., all \camtwo{its} entries are in \emph{occupied} state, \myName{} cannot keep track \camtwo{of information for a new synchronization variable} \camone{in} \myTableShort{}. \camone{We use the main memory as a fallback solution for \camtwo{such} \myTableShort{} overflow (Section~\ref{Overflowbl}). The \myEngineShort{} keeps track \camtwo{of} \emph{which} synchronization variables are currently serviced via main memory: similar to MiSAR~\cite{Liang2015MISAR}, we include \camtwo{a small set of counters (\emph{indexing counters}), 256 in current implementation,} indexed by the \camone{least significant bits} of the address of a synchronization variable, as extracted from the message that arrives \camthree{at} \camtwo{an} \myEngineShort{}.} When an \myEngineShort{} receives a message \camone{with acquire-type semantics for} a synchronization variable and there is no corresponding entry in the \emph{fully-occupied} \myTableShort{}, the indexing counter for that synchronization variable increases. When an \myEngineShort{} receives a message \camone{with release-type semantics for} a synchronization variable that is currently serviced using \camone{main} memory, the corresponding indexing counter decreases. \camone{A synchronization variable is currently serviced via main memory, when the corresponding indexing counter is larger than zero.} Note that different variables may alias to the same indexing counter. \camone{This aliasing does not affect correctness, but it does affect performance, \camtwo{since} a variable may unnecessarily be serviced via main memory, while the \myTableShort{} is \emph{not} full.}

\subsubsection{Control Flow in \myEngineShort{}} 
\begin{figure}[H]
  \vspace{2pt}
  \centering
  \includegraphics[width=1.02\columnwidth]{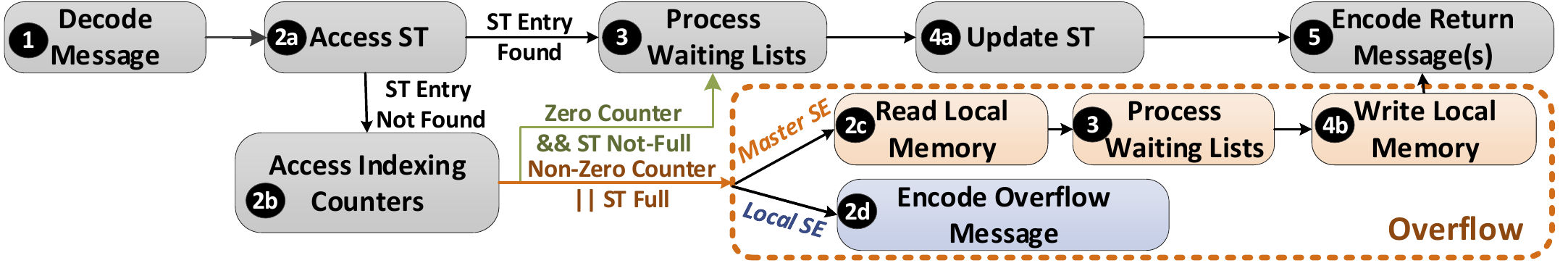}
  \caption{Control \camtwo{f}low in \myEngineShort{}.}
  \label{fig:control_flow}
\end{figure}

Figure~\ref{fig:control_flow} \camone{describes the control flow in \myEngineShort{}. When an \myEngineShort{} receives a message, it decodes the message \rectangled{1} and accesses the \myTableShort{} \rectangled{\hspace{1pt}2a\hspace{1.2pt}}. If there is an \myTableShort{} entry for the specific variable (depending on its address), the \myEngineShort{} processes the waiting lists \rectangled{3}, updates the \myTableShort{} \rectangled{\hspace{1pt}4a\hspace{1.2pt}}, and encodes return message(s) \rectangled{5}, if needed. 
If there is \emph{not} an \myTableShort{} entry for the specific variable, the \myEngineShort{} checks the value of the corresponding indexing counter \rectangled{\hspace{1pt}2b\hspace{1.2pt}}: (i) if the indexing counter is zero \emph{and} the \myTableShort{} is not full, the \myEngineShort{} reserves a \emph{new} \myTableShort{} entry and continues with step \rectangled{3}, otherwise (ii) if the indexing counter is larger than zero \emph{or} the \myTableShort{} is full, \camtwo{there is an overflow}. In that case, if the \myEngineShort{} is the \masterSE{} for the specific variable, it reads the synchronization variable from \emph{local} memory arrays \rectangled{\hspace{1pt}2c\hspace{1.2pt}}, processes the waiting lists \rectangled{3}, updates the variable in main memory \rectangled{\hspace{1pt}4b\hspace{1.2pt}}, and encodes return message(s) \rectangled{5}, if needed. If the \myEngineShort{} is \emph{not} the \masterSE{} for the specific variable, it encodes an \emph{overflow} message to the \masterSE{} \rectangled{\hspace{1pt}2d\hspace{1.2pt}} to handle overflow.
}

\subsection{Overflow Management}\label{Overflowbl} 

\myName{} integrates a hardware-only overflow management scheme that provides very modest performance degradation (Section~\ref{OverflowEvalbl}) and \camone{is programmer-transparent.} \camtwo{To handle \myTableShort{} overflow cases, we need to address two issues: (i) where to keep track of required information to coordinate synchronization, and (ii) how to coordinate \myTableShort{} overflow cases between \myEngineShort{}s. For the former issue, we design a generic structure allocated in main memory. 
For the latter issue, we propose a hierarchical \emph{overflow} communication protocol between \myEngineShort{}s.}

\subsubsection{\camtwo{\myName{}'s Synchronization Variable}}

\begin{figure}[t]
  \centering
  \includegraphics[scale=0.195]{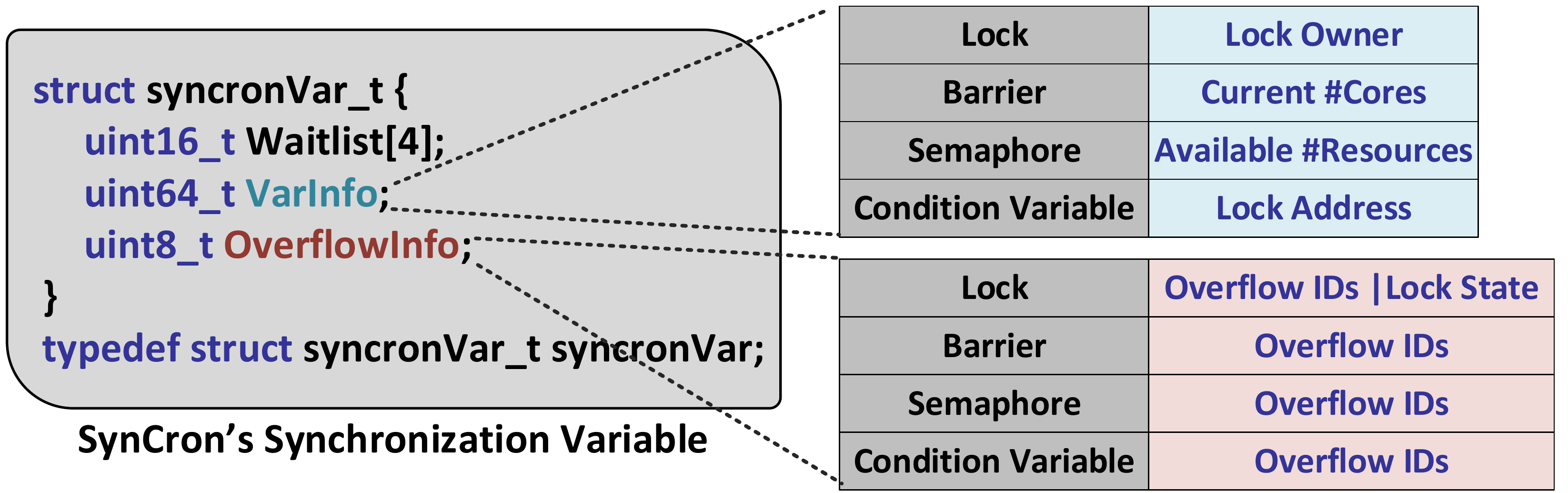}
  \caption{\camthree{Synchronization \camtwo{v}ariable of \myName{} (\emph{\mySyncVar{}})}.}
  \label{fig:syncVar}
  \vspace{-7pt}
\end{figure}

We design a generic structure (Figure~\ref{fig:syncVar}), 
called \emph{\mySyncVar{}}, which is used to coordinate synchronization for all supported primitives in \myTableShort{} overflow cases. \emph{\mySyncVar{}} is defined in the driver of the NDP system, which handles the allocation of the synchronization variables: \camtwo{programmers use \emph{create\_syncvar()} 
(Table~\ref{tab:interface}) to create a \emph{new} synchronization variable, the driver allocates the bytes needed for \emph{\mySyncVar{}} in main memory, and returns an opaque pointer that points to the address of the variable. Programmers should not de-reference \camthree{the} opaque pointer and its content can \emph{only} be accessed via \myName{}'s API (Table~\ref{tab:interface}).
}

\emph{\mySyncVar{}} structure includes one waiting list for each 
\camtwo{\myEngineShort{} of the system}, which has one bit for each NDP core within the NDP unit, and two additional fields (\emph{VarInfo, OverflowInfo}) \camtwo{needed to hierarchically handle \myTableShort{} overflows for all 
primitives.}


\subsubsection{\camtwo{Communication Protocol between \myEngineShort{}s}} 
To ensure correctness, \camtwo{\emph{only} the \masterSE{} updates the \emph{\mySyncVar{}} variable: in \myTableShort{} overflow, the SPU of the \masterSE{} issues read or write requests to its local memory to \emph{globally} coordinate synchronization \camtracy{via the} \emph{\mySyncVar{}} variable. In our proposed hierarchical design,} there are two overflow scenarios: (i) the \myTableShort{} of the \masterSE{} overflows, and (ii) the \myTableShort{} of a local \myEngineShort{} overflows \camtwo{or \myTableShort{}s of multiple local \myEngineShort{}s overflow}.

\noindent\textbf{The \myTableShort{} of the \masterSE{} overflows.} 
\camtwo{The other \myEngineShort{}s of the system have \emph{not} overflowed for a specific synchronization variable. Thus, they can still directly buffer this  variable in their local \myTableShort{}s, and serve their local cores 
\camtracy{themselves}, implementing a hierarchical (two-level) communication with \masterSE{}. 
The \masterSE{} receives \emph{global} messages from \myEngineShort{}s, and serves a local \myEngineShort{} of an NDP unit using \emph{all} bits in the waiting list of the \emph{\mySyncVar{}} variable associated with that local \myEngineShort{}. Specifically, when it receives a \emph{global} acquire-type message from a local \myEngineShort{}, it sets \emph{all} bits in the corresponding waiting list of the \emph{\mySyncVar{}} variable. When it receives a \emph{global} release-type message from a local \myEngineShort{}, it resets \emph{all} bits in the corresponding waiting list of the \emph{\mySyncVar{}} variable. }

\noindent\textbf{The \myTableShort{} of a local \myEngineShort{} overflows.}
In this scenario,
\camtwo{there are local \myEngineShort{}s that have overflowed for a specific variable, and local \myEngineShort{}s that have \emph{not} overflowed. Without loss of generality, we assume that only one \myEngineShort{} of the system has overflowed. \textbf{The local \myEngineShort{}s that have \emph{not} overflowed} serve their local cores 
\camtracy{themselves} via their \myTableShort{}s, implementing a hierarchical (two-level) communication with \masterSE{}. When the \masterSE{} receives a \emph{global} message from a local \myEngineShort{} (\camtracy{that} has \emph{not} overflowed), it (i) sets (or resets) \emph{all bits} in the waiting list of the \emph{\mySyncVar{}} variable associated with that \myEngineShort{}, and (ii) responds with a \emph{global} message to the local \myEngineShort{}, if needed.

\textbf{The overflowed \myEngineShort{}} needs to notify the \masterSE{} to handle \emph{local} synchronization requests of NDP cores located at \emph{another} NDP unit via main memory. We design \emph{overflow} message \camtwo{opcodes} (Table~\ref{tab:opcodes}) to be sent from the local overflowed \myEngineShort{} to the \masterSE{} and back. The overflowed \myEngineShort{} 
re-directs all messages (sent from its local NDP cores) for \camsix{a specific} variable to the \masterSE{} using the \emph{overflow} message opcodes, and both the overflowed \myEngineShort{} and the \masterSE{} increase their corresponding indexing counters to indicate that this variable is currently serviced via memory. When the \masterSE{} receives an \emph{overflow} message, it (i) sets (or resets) in the waiting list (associated with the overflowed \myEngineShort{}) of the \emph{\mySyncVar{}} variable, the bit that corresponds to the \emph{local ID} of the NDP core within the NDP unit, (ii) sets (or resets) in the \emph{OverflowInfo} field of the \emph{\mySyncVar{}} variable the bit that corresponds to the \emph{global ID} of the overflowed \myEngineShort{} to keep track of \emph{which} \myEngineShort{} (or \myEngineShort{}s) of the system has overflowed, and (iii) responds with an \emph{overflow} message to that \myEngineShort{}, if needed. The \emph{local ID} of the NDP core, and the \emph{global ID} of the overflowed \myEngineShort{} are encoded in the \emph{CoreID} field of the message (Figure~\ref{fig:msgencoding}). \camfour{When all bits in the waiting lists of the \emph{\mySyncVar{}} variable become zero (upon receiving a  release-type message), the \masterSE{} decrements the corresponding indexing counter. Then, it 
\camthree{sends a \emph{decrease\_index\_counter} message (Table~\ref{tab:opcodes}) to the overflowed \myEngineShort{} (\camfour{based on} the set bit \camfive{that is} tracked in the \emph{OverflowInfo} field), which decrements its corresponding indexing counter.}}



}

\subsection{\camtwo{\myName{} Enhancements}}

\subsubsection{\camone{\emph{RMW} Operations}}
It \camthree{is straightforward to extend \myName{}} to support simple \camone{atomic} \emph{rmw} operations inside \camone{the} \myEngineShort{} (by adding a lightweight ALU). \camtwo{The \masterSE{} could be responsible \camthree{for executing} atomic \emph{rmw} operations on a variable depending on its address.} We leave that for future work.

\subsubsection{Lock \camone{F}airness}
When local cores of an NDP unit repeatedly request a lock from their local \myEngineShort{}, the \myEngineShort{} repeatedly grants the lock within its unit, \camone{potentially causing unfairness and delay} to other NDP units. To prevent this, an extra field of a \camone{local grant} counter could be added to the \myTableShort{} entry. The counter increases every time the \myEngineShort{} grants the lock to a local core. If the counter exceeds a predefined threshold, then when the \myEngineShort{} receives a lock release, it transfers the lock to another \camtwo{\myEngineShort{}} \camtwo{(assuming other \myEngineShort{}s request the lock)}. The host OS or the user could dynamically set this threshold \camone{via a dedicated register. We leave the exploration of \camtwo{such fairness mechanisms to} future work.}

\subsection{Comparison with Prior Work}
\myName{}'s design shares some of its design concepts 
with SSB~\cite{Zhu2007SSB}, LCU~\cite{Vallejo2010Architectural}, and MiSAR~\cite{Liang2015MISAR}. However, \myName{} is more general\camtwo{, supporting the four most widely-used synchronization primitives,} and easy-to-use \camtwo{thanks} to its \camtwo{high-level} programming interface.

Table~\ref{tab:comparison} qualitatively compares \myName{} with these schemes. SSB and LCU support only lock semantics, thus they introduce two \emph{ISA extensions} for a simple lock. MiSAR introduces seven ISA extensions to support three primitives and handle overflow scenarios. \myName{} includes two ISA extensions for \camtwo{four} \emph{supported primitives}. A \emph{spin-wait approach} performs consecutive synchronization retries, typically incurring high energy consumption. A \emph{direct notification} scheme sends a direct message to only one waiting core when the synchronization variable becomes available, minimizing the traffic involved upon a release operation. \camtwo{SSB, LCU and MiSAR are tailored for \emph{uniform} memory systems. In contrast, \myName{} is the \emph{only} hardware synchronization mechanism that targets NDP systems as well as \emph{non-uniform} memory systems. }

SSB and LCU handle \emph{overflow} \camone{in hardware synchronization resources} using a pre-allocated table in \camtwo{main} memory, and if it overflows, they switch to software exception handlers \camtwo{(handled by \camthree{the} programmer), which} typically incur large overheads \camone{(due to OS intervention)} when overflows \camone{happen at a non-negligible frequency}. \camtwo{To avoid falling back to main memory,} which has high latency, \camtwo{and using expensive software exception handlers,} MiSAR requires the programmer to handle overflow scenarios using alternative \camone{software} synchronization \camone{libraries} \camthree{(e.g., pthread library provided by the OS).}
\camtwo{This approach can provide performance benefits in CPU systems, since alternative synchronization solutions can exploit low-cost accesses to caches and hardware cache coherence. However, in NDP systems alternative solutions would by default use main memory due to the absence of shared caches and hardware cache coherence support. Moreover, when} overflow occurs, \camone{MiSAR's accelerator} sends abort messages to all participating CPU cores notifying them to use the alternative \camtwo{solution, and when} \camone{the cores finish synchronizing via \camthree{the} alternative solution}, they notify \camone{MiSAR's} accelerator \camone{to switch \camtwo{back} to hardware synchronization}. \camtwo{This scheme introduces additional hardware/ISA complexity, and communication between the cores and the accelerator, thus \camthree{incurring} high network traffic and communication costs, as we show in Section~\ref{OverflowEvalbl}.} \camone{In contrast}, \myName{} directly falls back to memory \camtwo{via a \camthree{fully-integrated} hardware-only overflow scheme, which} provides graceful performance degradation (Section~\ref{OverflowEvalbl}), and \camtwo{is completely transparent to the programmer: programmers \camone{\emph{only}} use \myName{}'s high-level API, similarly to how software libraries are in charge of synchronization.}

\vspace{2pt}
\renewcommand{\arraystretch}{1.2}
\begin{table}[H]
  \begin{minipage}{.48\textwidth}
  \resizebox{\textwidth}{!}{
    \begin{tabular}{ l c c c c} 
\toprule
   & SSB~\cite{Zhu2007SSB} &
  LCU~\cite{Vallejo2010Architectural} &
  MiSAR~\cite{Liang2015MISAR} &
  \textbf{SynCron} \\
  \midrule
  \midrule
   
 Supported Primitives & 
  1 &
  1 &
  3 &
  \textbf{4} \\   \hline

  ISA Extensions &
  2 &
  2 &
  7 &
  \textbf{2} \\ \hline

  Spin-Wait Approach &
  yes &
  yes &
  no & 
  \textbf{no} \\ \hline
  
  Direct Notification &
  no &
  yes &
  yes &
  \textbf{yes} \\ \hline

  Target System &
  uniform &
  uniform &
  uniform &
  \textbf{non-uniform} \\ \hline
  
  Overflow &
  partially  &
  partially  &
  handled by  &
  \textbf{fully} \\ 
  
  Management &
  integrated &
  integrated &
  programmer &
  \textbf{integrated} \\ 
  

  \bottomrule
    \end{tabular}
    }
  \end{minipage}
  \caption{\label{tab:comparison}Comparison of \myName{} with prior mechanisms.}
  \vspace{-1pt}
\end{table}
\renewcommand{\arraystretch}{1}

\subsection{\camtwo{Use} of \myName{} in Conventional Systems}
The baseline NDP architecture \camtwo{~\cite{Ahn2015Scalable,Zhang2018GraphP,Youwei2019GraphQ,Tsai2018Adaptive,Gao2015Practical}} we assume in this work shares key design principles with \camone{conventional NUMA} systems. However, unlike NDP systems, NUMA CPU systems (i) have \camone{a shared level of cache (within a NUMA socket and/or across NUMA sockets)}, \camtwo{(ii) run multiple multi-threaded applications, i.e., a high number \camfive{of} software threads executed \camone{in} hardware thread \camone{contexts}, and \camfour{(iii)} the OS migrates \camfive{software} threads between hardware thread contexts to improve system performance.} Therefore, although \myName{} could be implemented in \camone{such commodity systems}, \camtwo{our} proposed hardware design \camfive{would need extensions. First, \myName{} could} exploit the low-cost accesses to \emph{shared} caches \camtwo{in conventional CPUs}, e.g., including an additional level in \myName{}'s hierarchical design to use \camtwo{the} shared cache for efficient synchronization within a NUMA socket, \camtwo{and/or} handling overflow scenarios by falling back to \camsix{the low-latency} cache instead of main memory. 
\camfour{Second, \myName{} needs} to support \camfive{use cases (ii) and (iii) listed above} in \camone{such} systems, i.e., including larger \myTableShort{}s and waiting lists to satisfy the needs of multiple multithreaded applications, handling the OS thread migration scenarios \camtwo{across} \camone{hardware thread contexts, and handling multiple synchronization requests sent from different software threads with the same hardware ID to \myEngineShort{}s, when \camthree{different} software threads are executed \camthree{on} the same hardware thread context.} We leave the \camtwo{optimization} of \myName{}'s design \camtwo{for} conventional systems to future work.

\setstretch{0.822}
\section{\camthree{Methodology}}\label{Methodologybl}

\camthree{\noindent\textbf{Simulation Methodology.}} We use an in-house simulator that integrates ZSim~\cite{Sanchez2013Zsim} and Ramulator~\cite{kim2015ramulator}. We model 4 NDP units (Table ~\ref{table:zsim_parameters}), each with 16 in-order cores. The cores issue a memory operation after the previous one has completed, i.e., there are no overlapping operations issued by the same core. Any write operation \camone{is} completed (and the latency is accounted for in our simulations) before executing the next instruction. To ensure memory consistency, compiler support~\cite{Rutgers2013Portable} guarantees that there is no reordering around the \emph{sync} instructions and a read is inserted after a write inside a critical section.

\begin{table}[H]
    \vspace{2pt}
    \hspace{-5pt}
    \resizebox{1\columnwidth}{!}{%
    \begin{tabular}{l l}
    \toprule
    \textbf{NDP Cores} & 16 in-order cores @2.5~GHz per NDP unit
    \\
    \midrule
    \textbf{L1 Data + Inst. Cache} & private, 16KB, 2-way, 4-cycle;  64 B line; 23/47 pJ per hit/miss~\cite{Muralimanohar2007Optimizing}  \\
    \midrule
    \textbf{NDP Unit} & buffered crossbar network with packet flow control; 1-cycle arbiter; \\
    \camone{\textbf{Local Network}}  & 1-cycle per hop~\cite{Agarwal2009garnet}; 0.4 pJ/bit per hop~\cite{Wolkotte2005Energy}; \\
    & M/D/1 model~\cite{Narayan2015} for queueing latency; \\ 
    \midrule
    \multirow{2}{*}{\textbf{DRAM HBM}} & 4 stacks; 4GB HBM 1.0~\cite{HBM_old,Lee2016Simultaneous}; 500MHz with 8 channels; \\
    & nRCDR/nRCDW/nRAS/nWR 7/6/17/8 \camthree{ns}~\cite{kim2015ramulator,Ghose2019Demystifying}; 7 pJ/bit~\cite{Mingyu2019Alleviating} \\ 
    \midrule
    \multirow{2}{*}{\textbf{DRAM HMC}} & 4 stacks; 4GB HMC 2.1; 1250MHz; 32 vaults per stack;  \\
    & nRCD/nRAS/nWR 17/34/19 \camthree{ns}~\cite{Ghose2019Demystifying,kim2015ramulator} \\
    \midrule
    \multirow{2}{*}{\textbf{DRAM DDR4}} & 4 DIMMs; 4GB \camone{each DIMM} DDR4 2400MHz; \\
    & nRCD/nRAS/nWR 16/39/18 \camthree{ns}~\cite{Ghose2019Demystifying,kim2015ramulator} \\
    \midrule
    \textbf{Interconnection Links} & 12.8GB/s per direction; 40 ns per cache line;  \\
    \camone{\textbf{Across NDP Units}} & 20-cycle; 4 pJ/bit\\
    \midrule
    \multirow{1}{*}{\textbf{Synchronization}} & SPU @1GHz clock frequency~\cite{shao2016aladdin}; \camone{8$\times$ 64-bit registers;} \\
    \textbf{Engine} & \camone{\camthree{b}uffer: 280B;} \myTableShort{}: 1192B, 64 entries, 1-cycle~\cite{Muralimanohar2007Optimizing}; \\
    & \camthree{i}ndexing \camthree{c}ounters: 2304B, 256 entries (8 LSB of the address), 2-cycle~\cite{Muralimanohar2007Optimizing} \\
    \bottomrule
    \end{tabular}
 }
 \caption{Configuration of our simulated system.}
\label{table:zsim_parameters}
\vspace{-4pt}
\end{table}

We evaluate three NDP configurations for different memory technologies, namely 2D, 2.5D, 3D NDP. The 2D NDP configuration \camtracy{uses a} DDR4 memory model and resembles recent 2D NDP systems~\cite{lavenier2016dna,upmem,devaux2019,gomezluna2021upmem}. In the 2.5D NDP configuration, each compute die of NDP units (16 NDP cores) is connected to an HBM stack via an interposer, \camtracy{similar} to current GPUs~\cite{mojumder2018profiling,designontap} and FPGAs~\cite{ultrascale,Singh2020NERO}. For the 3D NDP configuration, we use \camone{the} HMC memory model, \camone{where} the compute die of \camone{the} NDP unit \camone{is} located in the logic layer of the memory stack, as in prior works~\cite{Youwei2019GraphQ,Zhang2018GraphP,Ahn2015Scalable,Boroumand2018Google}. \camone{Due to space limitations, we present detailed evaluation results for the 2.5D NDP configuration, and provide a sensitivity study for the different NDP configurations in Section~\ref{MemoryTechnologies}.}

We model a crossbar network within \camone{each} NDP unit, simulating queuing latency using \camone{the} M/D/1 model~\cite{Narayan2015}. 
We count in ZSim-Ramulator all events for caches, i.e., number of hits/misses, network, i.e., number of bits transferred inside/across NDP units, and memory, i.e., number of total memory accesses, and use CACTI~\cite{Muralimanohar2007Optimizing} 
and parameters reported in prior \camtwo{works}~\cite{Mingyu2019Alleviating,Wolkotte2005Energy,Tsai2018Adaptive} to calculate energy. \camtwo{To estimate the latency in \myEngineShort{}, we use CACTI for \myTableShort{} and indexing counters, and Aladdin~\cite{shao2016aladdin} for the SPU with 1GHz at 40nm. Each message is served in 12 cycles, \camone{corresponding to} the message (barrier\_depart\_global) \camone{that takes} the longest time. }


\noindent\textbf{\camthree{Workloads.}} We evaluate workloads with both \camone{(i)} coarse-\camtwo{grained} synchronization, i.e., including \camone{only} a few synchronization variables to protect shared data, \camone{leading to} cores highly \camone{contending for} them (\camtracy{\emph{high-contention}}), and \camone{(ii)} fine-\camtwo{grained} synchronization, i.e., including a \camone{large} number of synchronization variables, each of them protecting a small granularity of shared data, \camone{leading to} cores not frequently \camone{contending for} the same variables at the same time (\camtracy{\emph{low-contention}}). We use the term \emph{synchronization intensity} to refer to the ratio of synchronization \camone{operations} over \camtwo{other} computation \camtwo{in} the workload. As this ratio increases, synchronization latency affects the total execution time of the workload \camone{more}. 

We study three classes of applications (Table~\ref{tab:workloads}), \camone{all} well suited for NDP. \camtwo{First, we evaluate pointer chasing workloads, i.e., lock-based concurrent data structures from the ASCYLIB library~\cite{David2015ASCYLIB}, used as key-value sets.} In ASCYLIB's Binary Search Tree (BST)~\cite{Drachsler2014Practical}, the 
lock memory requests are only 0.1\% of the total memory requests, \camone{so} we also evaluate an external fine-grained locking BST from~\cite{Siakavaras2017Combining}. Data structures are initialized with a fixed size and statically partitioned across NDP units, except for BSTs, which are distributed randomly. In these benchmarks, each core performs a fixed number of operations. We use lookup operations for data structures that support it, deletion for the rest, and push and pop operations for stack and queue. 
Second, we evaluate graph applications with fine-\camtwo{grained} synchronization \camtwo{from Crono}~\cite{Ahmad2015CRONO,hong2014simplifying} (push version), where the output array has read-write data. \camthree{All real-world graphs~\cite{davis2011florida} used are undirected} and \camone{statically} partitioned across NDP units, \camone{where the} vertex data \camone{is} equally distributed across \camthree{cores}. Third, we evaluate time series analysis~\cite{TL17}, using SCRIMP, and \emph{real} data sets from Matrix Profile~\cite{MPROFILEI}. We replicate the input data in each NDP unit and partition the output array (read-write data) across \camone{NDP} units.

\begin{table}[!htb]
\begin{minipage}{.4\textwidth}
   \hspace{17pt}
   \resizebox{\textwidth}{!}{
    \begin{tabular}{l l} 
    \toprule
    \textbf{Data Structure} &  \textbf{Configuration} \\ 
    \midrule
    \midrule
    Stack~\cite{David2015ASCYLIB} & 100K - 100\% push \\
    Queue~\cite{Michael1996Simple,David2015ASCYLIB} & 100K - 100\% pop \\
    Array Map~\cite{David2015ASCYLIB,Guerraoui2016Optimistic} & 10 - 100\% lookup \\ 
    Priority Queue~\cite{David2015ASCYLIB,Pugh1990Concurrent,Alistarh2015Spraylist} & 20K - 100\% deleteMin \\ 
    Skip List~\cite{David2015ASCYLIB,Pugh1990Concurrent} & 5K - 100\% deletion\\ 
    Hash Table~\cite{David2015ASCYLIB,herlihy2008art} & 1K - 100\% lookup\\
    Linked List~\cite{David2015ASCYLIB,herlihy2008art} & 20K - 100\% lookup \\
    Binary Search Tree Fine-Grained (BST\_FG)~\cite{Siakavaras2017Combining}
    & 20K - 100\% lookup \\ 
    Binary Search Tree Drachsler (BST\_Drachsler)~\cite{David2015ASCYLIB,Drachsler2014Practical} & 10K - 100\% deletion \\
    \bottomrule
    \end{tabular}}
  \end{minipage}
  \begin{minipage}{.5\textwidth}
  \vspace{1pt}
  \hspace{-7pt}
  \resizebox{\textwidth}{!}{
    \begin{tabular}{l c c} 
    \toprule
    \textbf{Real Application} & \textbf{Locks} & \textbf{Barriers} \\ 
    \midrule
    \midrule
    Breadth First Search (\textbf{bfs})~\cite{Ahmad2015CRONO} & \checkmark & \checkmark \\ 
    Connected Components (\textbf{cc})~\cite{Ahmad2015CRONO} & \checkmark & \checkmark \\
    Single Source Shortest Paths (\textbf{sssp})~\cite{Ahmad2015CRONO} & \checkmark & \checkmark \\
    Pagerank (\textbf{pr})~\cite{Ahmad2015CRONO} & \checkmark & \checkmark \\
    Teenage Followers (\textbf{tf})~\cite{hong2014simplifying} & \checkmark  & - \\ 
    Triangle Counting (\textbf{\camone{tc}})~\cite{Ahmad2015CRONO}  & \checkmark  & \checkmark \\ 
    \midrule 
    Time Series Analysis (\textbf{ts})~\cite{MPROFILEI} & \checkmark & \checkmark \\
    \bottomrule
    \end{tabular}
    \hspace{1pt}
    \begin{tabular}{l c}
    \toprule
    \textbf{Real Application} & \textbf{Input Data Set} \\ 
    \midrule
    \midrule
     & wikipedia \\ 
     & -20051105 (\textbf{wk}) \\
   \textbf{bfs, cc, sssp,} & soc-LiveJournal1 (\textbf{sl})\\  
   \textbf{pr, tf, tc} & sx-stackoverflow (\textbf{sx}) \\ 
    & com-Orkut (\textbf{co}) \\ 
    \midrule 
    \multirow{2}{*}{\shortstack[l]{\textbf{ts}}} & air quality (\textbf{air})\\
    & energy consumption (\textbf{pow}) \\ 
    \bottomrule
    \end{tabular}
    }
  \end{minipage}
  \caption{\label{tab:workloads} Summary of all workloads used in our evaluation.}
  \vspace{-6pt}
\end{table}

\noindent\textbf{\camthree{Comparison Points.}} We compare \myName{} with three schemes: (i) \emph{Central}: a \mpsync{} scheme that supports all primitives by extending the barrier primitive of \camtwo{Tesseract}~\cite{Ahn2015Scalable}, i.e., one dedicated NDP core \camone{in the entire NDP system} acts as server and coordinates synchronization among all NDP cores of the system by issuing memory requests to synchronization variables \camone{via} its memory hierarchy, while the \camone{remaining} client cores communicate with it via hardware message-passing; (ii) \emph{Hier}: a hierarchical \mpsync{} scheme that supports all \camtracy{primitives, similar} to the barrier primitive of~\cite{Gao2015Practical} (or hierarchical lock of~\cite{Tang2019plock}), i.e., one NDP core per NDP unit acts as server and coordinates synchronization by issuing memory requests to synchronization variables \camone{via} its memory hierarchy (including caches), and communicates with other servers \camtwo{and local client cores (located at the same NDP unit with it)} via hardware message-passing; (iii) \emph{Ideal}: an ideal scheme with zero \camone{performance overhead} for synchronization. In our evaluation, each NDP core runs one thread. \camthree{For fair comparison, we use the same number of client cores, i.e., 15 per NDP unit, that execute the main workload \camtwo{for all schemes}. For synchronization, we add one \camfour{server core} for the entire system in \naive{}, one \camfour{server core} per NDP unit for \hier{}, and one \myEngineShort{} per NDP unit for \myName. For \myName{}, \camfour{we disable one core per NDP unit} to match the same number of client cores as the previous schemes.} Maintaining the same thread-level parallelism for executing the main kernel is \camtwo{consistent} with prior \camone{works} on message-passing synchronization~\cite{Tang2019plock,Liang2015MISAR}. 


\setlength{\intextsep}{3pt} 
\setstretch{0.848}
\captionsetup[figure]{aboveskip=0.2em, belowskip=0.em} 
\section{Evaluation}\label{Evaluationbl}

\setstretch{0.844}

\subsection{Performance}\label{Performancebl}

\subsubsection{Synchronization Primitives} 
\begin{figure}[H]
\hspace{-3pt}
  \includegraphics[width=0.99\columnwidth]{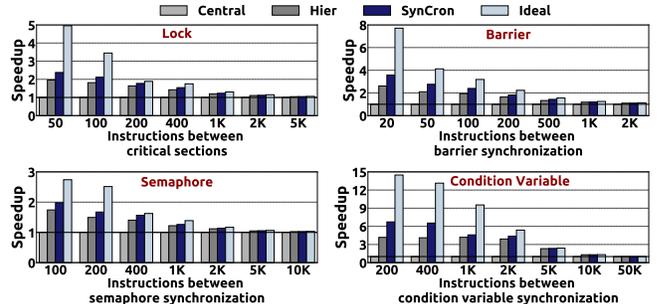}
  \caption{Speedup of different synchronization primitives.}
   \label{fig:mbench}
\end{figure}

Figure~\ref{fig:mbench} evaluates all supported primitives using 60 cores, varying the interval \camtwo{(in terms of instructions)} between two synchronization points. We devise simple benchmarks, where cores repeatedly request a single synchronization variable. For lock, \camone{the critical section is empty, i.e., it does not include any instruction}. For semaphore and condition variable, half of the cores execute sem\_wait/cond\_wait, while the rest \camtracy{execute} sem\_post/cond\_signal, respectively. \camone{As the interval between synchronization points becomes smaller}, \myName{}'s performance benefit \camone{increases}. For an interval of 200 instructions, \myName{} outperforms \naive{} and \hier{} by 3.05$\times$ and 1.40$\times$ respectively, averaged across all primitives. \camtwo{\myName{} outperforms \hier{} due to directly buffering synchronization variables \camthree{in low-latency} \myTableShort{}s, and achieves the highest benefits in the condition variable primitive (by 1.61$\times$), since this benchmark has higher synchronization intensity compared to the rest: cores coordinate for both the condition variable and the lock associated with it.} \camone{When \camtwo{the} interval between synchronization operations becomes larger}, synchronization requests become less dominant in the main workload, and thus all schemes perform similarly. Overall, \myName{} outperforms prior schemes for all different synchronization primitives.

\vspace{-2pt}

\setstretch{0.848}

\subsubsection{Pointer Chasing \camtwo{Data Structures}} 
\begin{figure}[H]
\hspace{-5pt}
  \begin{subfigure}[h]{\textwidth}
   \includegraphics[scale=0.144]{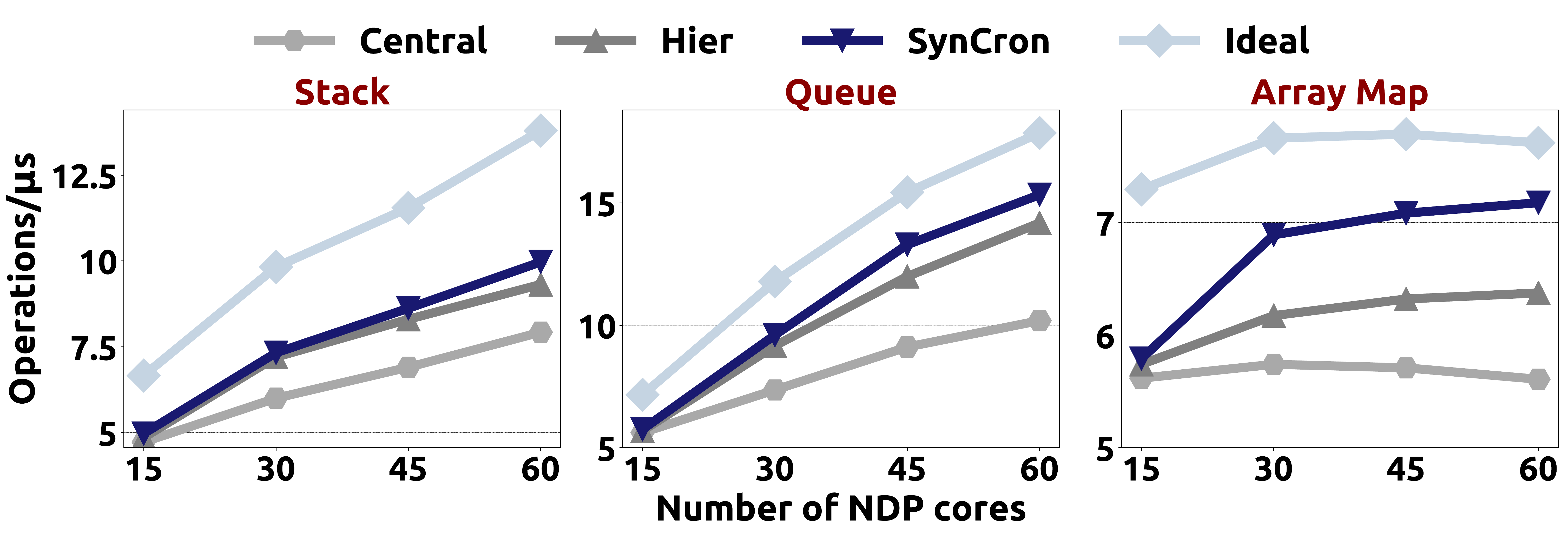}
  \end{subfigure}
  \begin{subfigure}[h]{\textwidth}
  \hspace{-4pt}
   \includegraphics[scale=0.148]{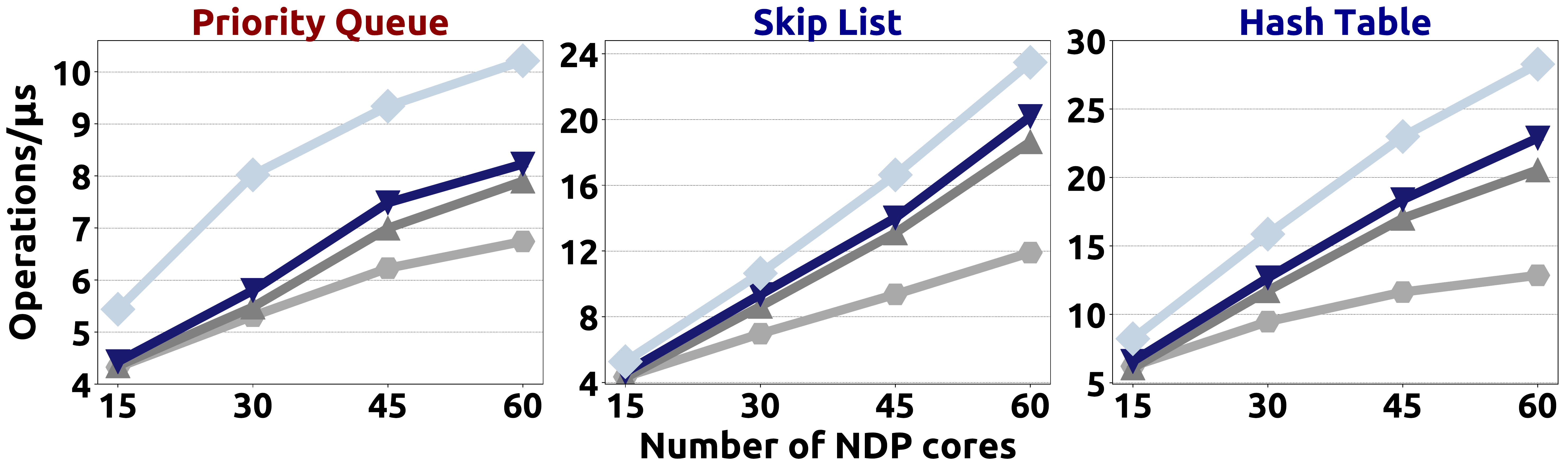}
  \end{subfigure} 
  \begin{subfigure}[h]{\textwidth}
  \hspace{-4pt}
   \includegraphics[scale=0.147]{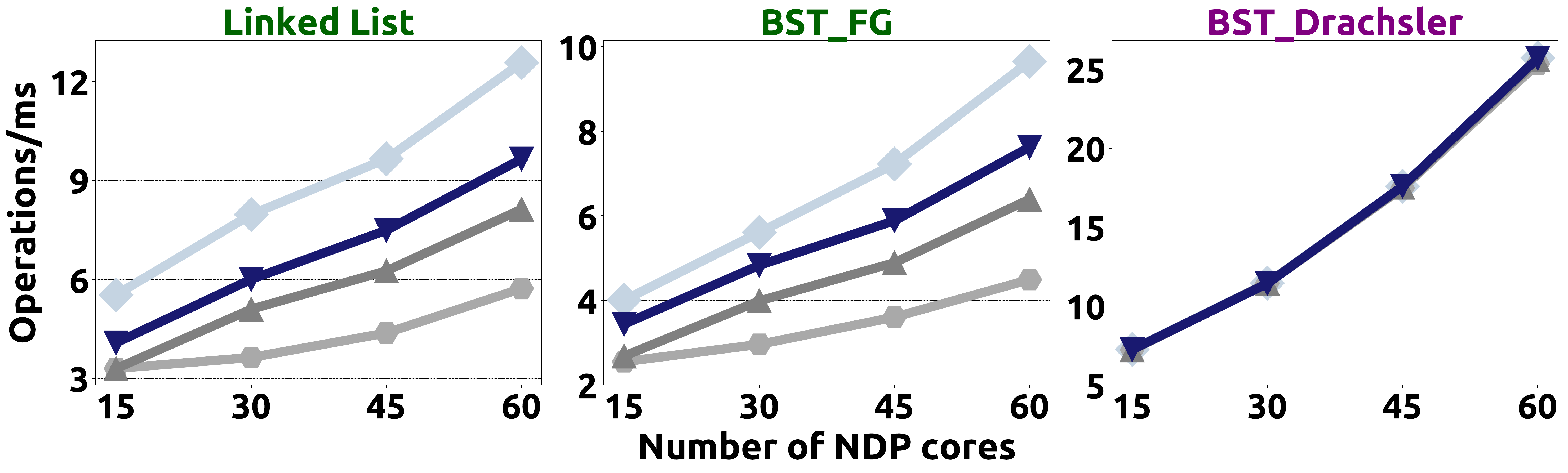}
  \end{subfigure}
  \caption{Throughput of pointer chasing using data structures.}
  \label{fig:ds_thr}
  \vspace{-4pt}
\end{figure}

Figure~\ref{fig:ds_thr} shows the throughput for all schemes in pointer chasing \camtwo{varying the NDP cores in steps of 15, \camtracy{each} time adding one NDP unit.}

\begin{figure*}[!b]
\vspace{3pt}
\centering
  \hspace{-5pt}
  \includegraphics[scale=0.033]{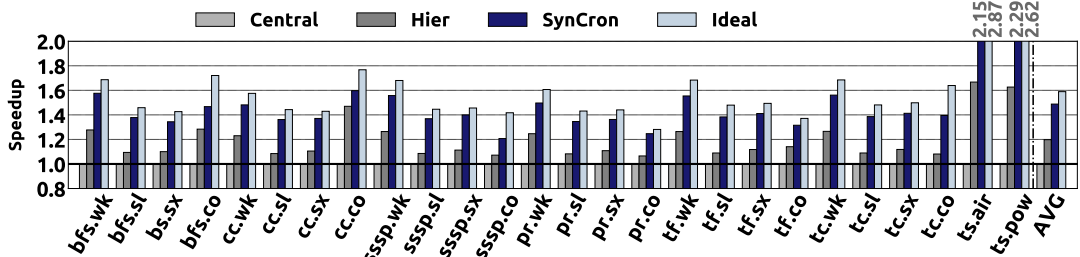}
  \caption{Speedup in real applications normalized to \naive{}.}
  \label{fig:speedup_real}
\end{figure*}

We observe four different patterns. First, \emph{stack}, \emph{queue}, \emph{array map}, and \emph{priority queue} incur \camtracy{high contention}, as all cores \camone{heavily contend} for a few variables. \emph{Array map} has the lowest scalability due to a larger critical section. In high-contention scenarios, hierarchical schemes (\hier{}, \myName{}) perform better by \camone{reducing} the expensive traffic across NDP units. \myName{} outperforms \hier{}, since the latency cost of using \myEngineShort{}s \camone{that} update \camone{small} \myTableShort{}s is lower than using NDP cores as servers \camone{that} update larger caches. Second, \emph{skip list} and \emph{hash table} incur \camtracy{medium contention}, as different cores may work on different parts of the data structure. \camtwo{For these data structures,} hierarchical schemes perform better, as they minimize the expensive traffic, and multiple server cores concurrently serve requests to their local memory. \myName{} retains most of the performance benefits of \ideal{}, incurring \camone{only} 19.9\% overhead with 60 cores, \camthree{and} outperforms \hier{} by 9.8\%. Third, \emph{linked list} and \emph{BST\_FG} exhibit \camtracy{low contention} and high synchronization demand, as each core requests multiple locks concurrently. These data structures cause \camtwo{higher} \camone{synchronization-related} traffic inside the network \camtwo{compared to \emph{skip list} and \emph{hash table}}, \camone{and} thus \myName{} \camtwo{further} outperforms \hier{} by 1.19$\times$ due to directly buffering synchronization variables \camone{in} \myTableShort{}s. Fourth, in \emph{BST\_Drachsler} lock requests constitute only 0.1\% of the total requests, and all schemes perform similarly.
Overall, \camone{we conclude that} \myName{} achieves higher throughput \emph{than prior mechanisms} under different scenarios with diverse conditions.

\subsubsection{Real \camone{A}pplications} 
Figure~\ref{fig:speedup_real} shows the performance of all schemes with real applications using all NDP units, normalized to \naive{}. \camone{Averaged across 26 application-input combinations,} \myName{} outperforms \naive{} by 1.47$\times$ and \hier{} by \camone{1.23$\times$}, and \camtwo{performs} \camone{within 9.5\% of} \ideal{}.

Our real applications \camseven{exhibit} \camtracy{low contention}, as two cores rarely \camone{contend} for the same synchronization variable, and high synchronization demand, as several synchronization variables are active during execution. We observe that \hier{} and \myName{} increase parallelism, because the \camtwo{per-NDP-unit} servers service different synchronization requests concurrently, and avoid remote synchronization messages across NDP units. Even though \hier{} \camone{performs 1.19$\times$ better} than \naive{}, \camone{on average}, \camtwo{its performance is still 1.33$\times$ worse than} \ideal{}. \myName{} \camone{provides} most of the performance benefits of \ideal{} (with only 9.5\% overhead on average), and outperforms \hier{} \camone{due to} directly buffering the synchronization variables \camone{in} \myTableShort{}s\camnine{, thereby} completely avoiding the memory accesses for synchronization requests. 
\camone{Specifically}, we find that \emph{time series analysis} has high synchronization intensity, since the ratio of synchronization over \camtwo{other} computation of the workload \camtwo{is} higher compared to graph workloads. For this application, \hier{} and \myName{} outperform \naive{} by 1.64$\times$ and 2.22$\times$, as \camone{they serve} multiple \camone{synchronization} requests concurrently. \myName{} further outperforms \hier{} by 1.35$\times$ due to directly buffering \camone{the synchronization variables in \myTableShort{}s}. We conclude that \myName{} performs best across \emph{all} real \camone{application-input combinations} and \camtwo{approaches} \camone{the} \ideal{} scheme \camone{with no synchronization overhead.}

\begin{figure}[H]
\vspace{1pt}
\centering
  \includegraphics[width=1.0\columnwidth]{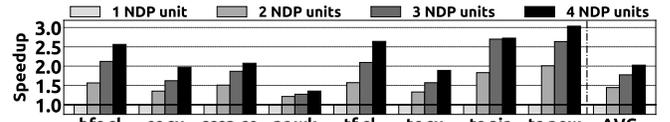}
  \caption{Scalability of real applications using \myName{}.}
  \label{fig:scalability}
\end{figure}

\noindent\textbf{Scalability.} 
Figure~\ref{fig:scalability} shows the scalability \camone{of real applications using \myName{} from 1 to 4 NDP units. \camtwo{Due to space limitations, we present a subset of our workloads, but we report average values for all 26 application-input combinations. This also applies for all figures presented henceforth.} Across all workloads, \myName{} enables performance scaling by at least 1.32$\times$, on average 2.03$\times$, and up to 3.03$\times$, when using 4 NDP units (60 NDP cores) over 1 NDP unit (15 NDP cores).}

\subsection{Energy Consumption}
\begin{figure}[t]
 \hspace{-5pt}
  \includegraphics[width=1.02\columnwidth]{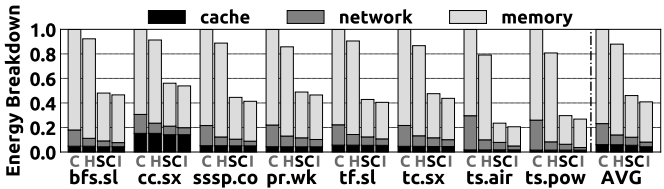}
  \caption{Energy breakdown in real applications for C: \naive{}, H: \hier{}, SC: \myName{} and I: \ideal{}.}
  \label{fig:energy}
  \vspace{-6pt}
\end{figure}

Figure~\ref{fig:energy} shows the energy breakdown for cache, network, and memory \camone{\camtwo{in} our real applications when} using all cores. \myName{} reduces the network and memory energy thanks to \camone{its} hierarchical design and direct buffering. On average, \myName{} reduces energy consumption by 2.22$\times$ over \naive{} \camone{and} 1.94$\times$ over \hier{}, and \camtwo{incurs only 6.2\% energy overhead over \ideal{}.}

We observe that 1) cache energy consumption constitutes a \camtwo{small} portion of the total energy, since these applications have irregular access patterns. NDP cores that act as servers for \naive{} and \hier{} increase the cache energy \camone{only} by 5.1\% and 4.8\% over \ideal{}. \camone{2)} \naive{} generates a larger amount of expensive traffic across NDP units compared to hierarchical schemes, resulting \camthree{in 2.68$\times$} \camfour{higher} network energy over \myName{}. \myName{} also \camtwo{has} less network energy (1.21$\times$) than \hier{}, because it avoids transferring synchronization variables from memory to \myEngineShort{}s due to directly buffering them. \camone{3)} \hier{} and \naive{} have approximately the same \camone{memory energy} consumption, \camone{because they issue} a similar number of requests to memory. \camone{In contrast}, \myName{}'s memory energy \camone{consumption is similar to that of} \ideal{}. We note that \myName{} provides \emph{higher} energy \camone{reductions} in applications with high synchronization intensity, such as time series analysis, since it avoids a \emph{higher} number of memory accesses for synchronization due to \camone{its} direct buffering \camtwo{capability}.

\subsection{Data Movement}
Figure~\ref{fig:traffic} shows \camone{normalized} data movement, i.e., bytes transferred between NDP cores and memory, for all schemes using \camtwo{four} NDP units. \camone{\myName{} reduces data movement across all workloads by 2.08$\times$ and 2.04$\times$ over \naive{} and \hier{}, respectively, \camone{on average,} and incurs \camtwo{only} 13.8\% \camtwo{more data movement than} \ideal{}.} \naive{} generates high data movement across NDP units, particularly when running time series analysis that has high synchronization intensity. \hier{} reduces the traffic across NDP units; however, it \camthree{may increase} the traffic inside \camthree{an NDP unit}, occasionally leading to slightly higher total data movement (e.g., \emph{ts.air}). This is because when an NDP core requests a synchronization variable that is physically located \camone{in} another NDP unit, it first sends a message inside the NDP unit to its local server, which in turns sends a message to the global server. \camone{In contrast}, \myName{} reduces the traffic \camtwo{inside \camthree{an} NDP unit due to \camthree{directly} buffering synchronization variables, and across NDP units due to its hierarchical design.
}

\begin{figure}[H]
  \hspace{-5pt}
  \includegraphics[width=1.02\columnwidth]{graphs/traffic.pdf}
  \caption{Data movement in real applications for C: \naive{}, H: \hier{}, SC: \myName{} and I: \ideal{}.}
  \label{fig:traffic}
  \vspace{1pt}
\end{figure}

\subsection{Non-Uniformity of NDP Systems}
\subsubsection{High \camtracy{\camone{C}ontention}} 
\begin{figure}[H]
  \centering
  \includegraphics[width=0.92\columnwidth]{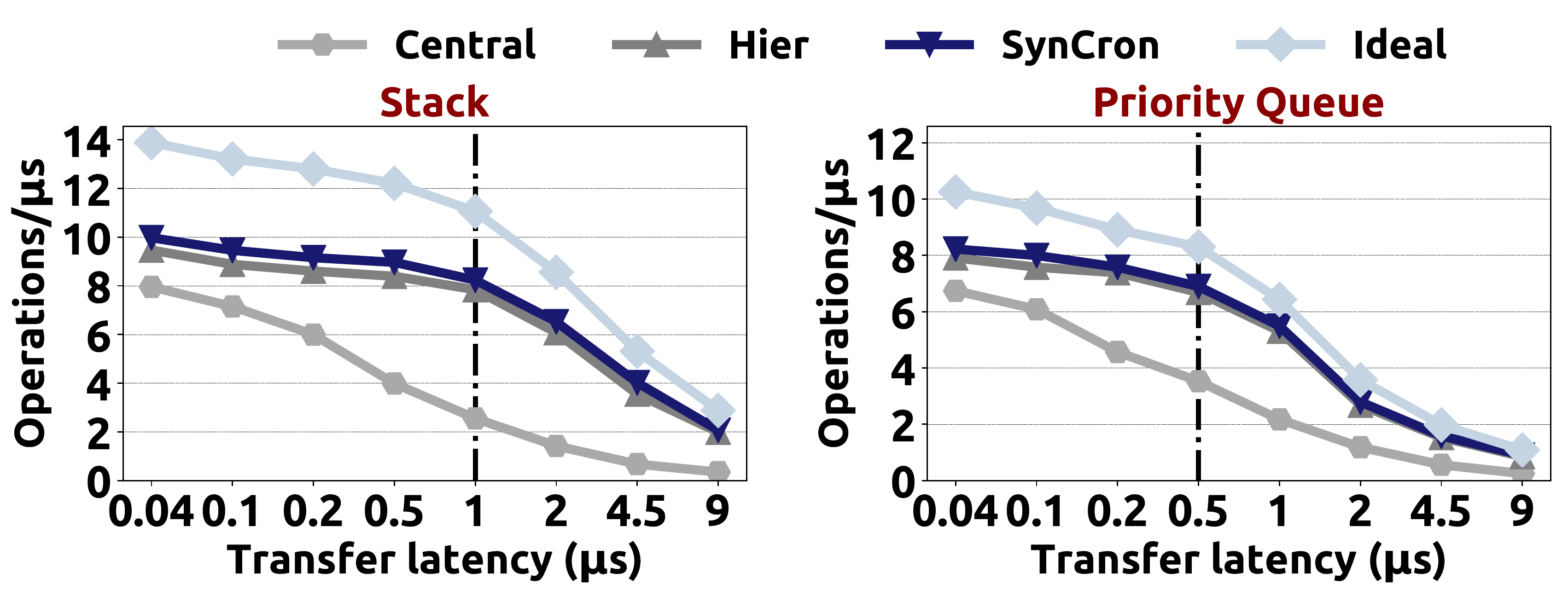}
  \vspace{-1pt}
  \caption{\camone{Performance sensitivity} to the transfer latency of the interconnection links used to connect the NDP units.}
  \label{fig:cds_numa}
  \vspace{-1pt}
\end{figure}

Hierarchical schemes provide high benefit \camtwo{under} \camtracy{high contention}, as they prioritize local requests inside \camone{each} NDP unit. We study their performance benefit in stack and priority queue (Figure~\ref{fig:cds_numa}) when varying the transfer latency of the interconnection links used across \camtwo{four} NDP units. \naive{} is significantly affected by the interconnect \camone{latency} across NDP units, as it is oblivious to the non-uniform nature of the NDP system. Observing \ideal{}, which reflects the actual behavior of the main workload, we notice that after a certain point (vertical line), the cost of remote memory accesses across NDP units become high \camtwo{enough to} dominate performance. \myName{} and \hier{} tend to follow the actual behavior of the workload, as local synchronization messages within NDP units are \camtwo{much} less expensive than remote messages of \naive{}. \myName{} outperforms \hier{} by 1.06$\times$ and 1.04$\times$ for stack and priority queue. \camone{We conclude that} \myName{} \camone{is the best at} hiding the latency of slow links across NDP units.

\vspace{-2pt}

\subsubsection{Low \camtracy{\camone{C}ontention}} 
\begin{figure}[H]
  \centering
  \includegraphics[width=0.93\columnwidth]{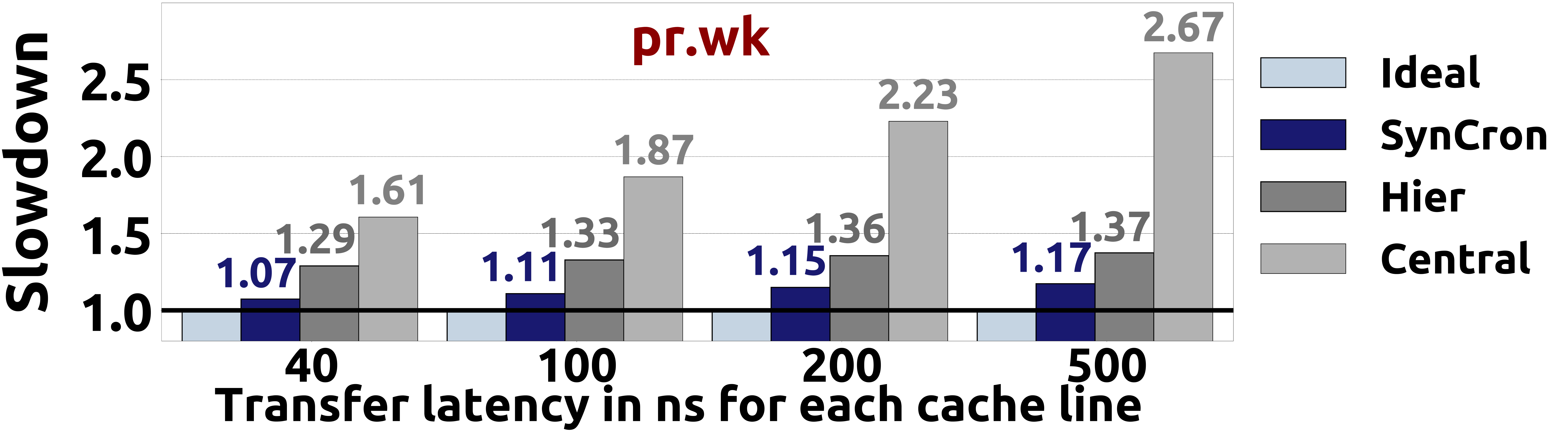}
  \vspace{-1pt}
  \caption{\camone{Performance sensitivity} to the transfer latency of the interconnection links \camnine{used to connect} the NDP units. \camone{All \camtwo{data is} normalized to \ideal{}} (\emph{lower is better}).}
  \label{fig:numa_pr}
\end{figure}

We also study the effect of interconnection links used across the NDP units \camone{in} a \camtracy{low-contention} graph application (Figure~\ref{fig:numa_pr}). Observing \ideal{}, with 500 \camthree{ns} transfer latency per cache line, \camone{we note that} the workload \camtwo{experiences} 2.46$\times$ slowdown over the default latency of 40 \camthree{ns}, as 24.1\% \camtwo{of its memory accesses are to remote NDP units.} As the transfer latency increases, \naive{} incurs significant slowdown over \ideal{}, since all NDP cores of the system communicate with one single server, generating expensive traffic across NDP units. \camone{In contrast}, the slowdown of hierarchical schemes over \ideal{} \camone{is smaller}, as \camone{these schemes} generate less remote traffic by distributing the synchronization requests across multiple local servers. \myName{} outperforms \hier{} due to \camone{its} direct buffering \camone{capabilities}. Overall, \myName{} outperforms prior high-performance schemes even \camone{when the} network delay \camtwo{across NDP units is large.}

\subsection{Memory Technologies}\label{MemoryTechnologies}
\begin{figure}[H]
\hspace{-4pt}
  \includegraphics[width=1.0\columnwidth]{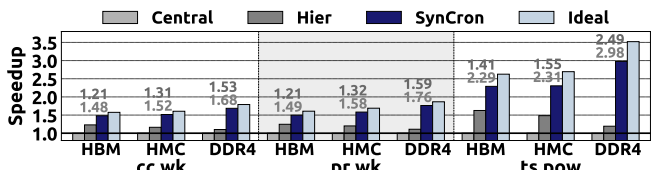}
  \vspace{-1pt}
  \caption{Speedup \camone{with} different memory technologies.}
  \label{fig:3D_memories}
\end{figure}

We study three memory technologies, which provide different memory access latencies \camone{and bandwidth}. We evaluate (i) 2.5D NDP using HBM, (ii) 3D NDP using HMC, and (iii) 2D NDP using DDR4. Figure~\ref{fig:3D_memories} shows the performance of all schemes normalized to \naive{} of each memory. The reported values show the speedup of \myName{} over \naive{} and \hier{}. \camthree{\myName{}'s benefit} is independent \camone{of} the memory used: its performance \camtwo{versus} \ideal{} only slightly varies ($\pm$1.4\%) across different memory technologies, \camtwo{since \myTableShort{}s never overflow.} Moreover, \myName{}'s \camthree{performance improvement} over prior schemes \camthree{increases} as the memory access latency becomes higher thanks to direct buffering, \camtwo{which avoids} expensive memory accesses for synchronization. \camthree{For example, in \emph{ts.pow}, \myName{} outperforms \hier{} by 1.41$\times$ and 2.49$\times$ with HBM and DDR4, respectively, as the latter incurs higher access latency. Overall, \myName{} is orthogonal to the memory \camtwo{technology} used.} 

\subsection{Effect of Data Placement}
\begin{figure}[h]
   \vspace{1pt}
   \begin{minipage}{\columnwidth}
  \includegraphics[width=\columnwidth]{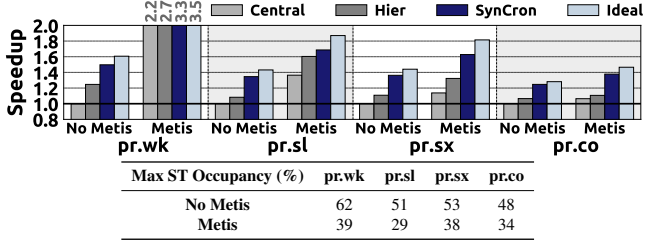}
  \end{minipage}
  \begin{minipage}{\columnwidth} \vspace{1pt}
  \centering
  \resizebox{0.64\columnwidth}{!}{%
    \begin{tabular}{c c c c c} 
    \toprule
    \textbf{Max \camone{\myTableShort{}} Occupancy (\%)} & \textbf{pr.wk} & \textbf{pr.sl} & \textbf{pr.sx} & \textbf{pr.co}  \\ 
    \midrule
    \textbf{No Metis} & 62 & 51 & 53 & 48 \\
    \textbf{Metis} & 39 & 29 & 38 & 34 \\
    \bottomrule
    \end{tabular}}
  \end{minipage}%
  \caption{\camone{Performance sensitivity} to a better graph \camthree{partitioning} \camone{and maximum \myTableShort{} occupancy of \myName{}.}}
  \label{fig:metis}
\end{figure}

Figure~\ref{fig:metis} evaluates the effect of better data placement on \myName{}'s benefits. We use Metis~\cite{Karypis1998Metis} to obtain a 4-way graph partitioning to minimize the crossing edges between the 4 NDP units. All \camone{data values} are normalized to \naive{} without Metis. \camone{For \myName{}, we define \myTableShort{} occupancy as the average fraction of ST entries that are occupied in each cycle.}

\camone{We make three observations. First,} \ideal{}, which reflects the actual behavior of the main kernel (i.e., with zero synchronization overhead), improves performance by 1.47$\times$ across the four graphs. \camone{Second,} with a better graph \camthree{partitioning}, \myName{} still outperforms both \naive{} and \hier{}. \camone{Third, we} find that \myTableShort{} occupancy \camtwo{is lower with a better graph \camthree{partitioning}}. When a local \myEngineShort{} receives a request for a synchronization variable of another NDP unit, \emph{both} the local \myEngineShort{} and the \masterSE{} reserve a new entry in their \myTableShort{}s. With a better graph \camthree{partitioning}, NDP cores send requests to their local \myEngineShort{}, which is also the \masterSE{} for the requested variable. Thus, \emph{only one} \myEngineShort{} of the system reserves a new entry, resulting in a lower \myTableShort{} occupancy. \camone{We conclude that}, with better \camthree{data placement} \myName{} still performs \camthree{the} best while achieving even lower \myTableShort{} occupancy.

\subsection{\myName{}'s Design Choices}\label{DesignChoices}
\vspace{1pt}

\subsubsection{Hierarchical Design}\label{Flat}
To demonstrate the effectiveness of \myName{}'s hierarchical design in non-uniform NDP systems, we compare it with \camfour{\myName{}'s \emph{flat} variant. Each core in 
\emph{flat} \emph{directly} sends all its synchronization requests to the \masterSE{} of each variable. In contrast, each core in \myName{} sends all its synchronization requests to the local \myEngineShort{}. If the local \myEngineShort{} is \emph{not} the \masterSE{} for the requested variable, the local \myEngineShort{} sends a message across NDP units to the \masterSE{}. 
}

We evaluate \camone{\camthree{three} synchronization scenarios: (i) low-contention and synchronization non-intensive (e.g., graph applications), (ii) low-contention and synchronization-intensive (e.g., time series analysis), and (iii) high-contention (e.g., queue data structure). 

\noindent\textbf{Low-contention and synchronization non-intensive.} 
\camfour{Figure~\ref{fig:hier_benefit_graphs} evaluates this scenario using several graph processing workloads with 40 \camthree{ns} \camthree{link} latency \camthree{between} NDP units. \myName{} is 1.1\% worse than \camfive{\emph{flat},} on average. We conclude that \myName{} performs only \emph{slightly} worse than \emph{flat} for low-contention and synchronization non-intensive scenarios.}

\begin{figure}[H]
  \vspace{1pt}
  \hspace{-4pt}
  \includegraphics[width=1.02\columnwidth]{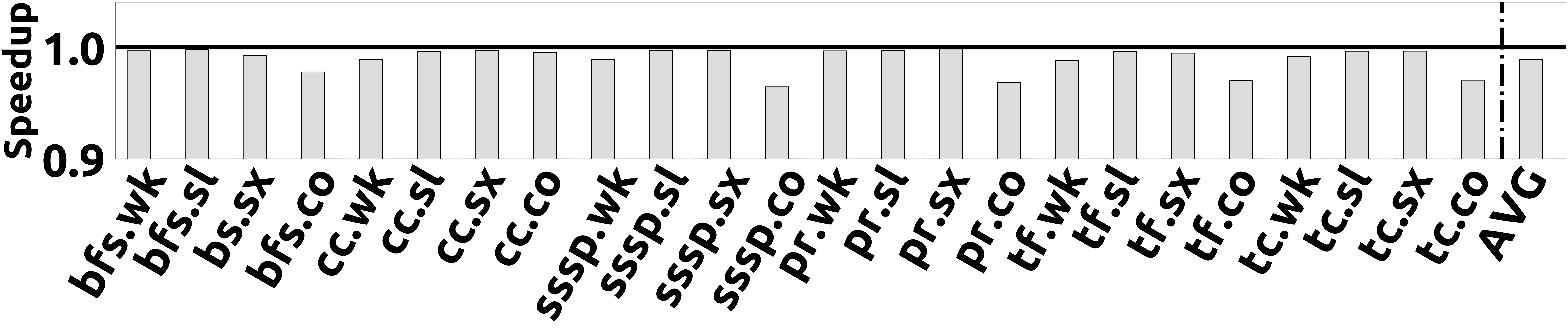}
  \caption{Speedup of \myName{} normalized to \emph{flat} \camthree{with 40 ns} \camthree{link} latency \camthree{between NDP units}, \camone{under} a low-contention \camone{and synchronization non-intensive scenario}.}
  \label{fig:hier_benefit_graphs}
 \vspace{-4pt}
\end{figure}

\noindent\textbf{Low-contention and synchronization-intensive.} 
\camfour{Figure~\ref{fig:hier_benefit}a evaluates this scenario using time series analysis with four different \camthree{link} latency values \camthree{between} NDP units.} \myName{} performs 7.3\% worse than \emph{flat} with a 40 ns inter-NDP-unit latency. With a 500 \camthree{ns} inter-NDP-unit latency, \myName{} performs \emph{only} 3.6\% worse than \emph{flat}, since remote traffic has a larger impact on the total execution time. We conclude that \myName{} performs modestly worse than \emph{flat}, and \myName{}'s slowdown decreases as non-uniformity, i.e., the latency \camthree{between} NDP units, increases.

\begin{figure}[H]
  \centering
  \includegraphics[scale=0.0264]{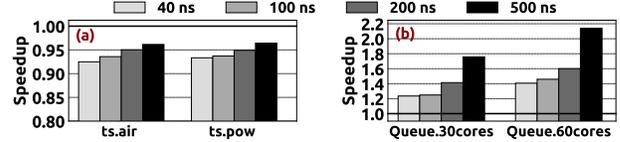}
  \caption{Speedup of \myName{} normalized to \emph{flat}, \camone{as we vary} the transfer latency of the interconnection links used to connect NDP units, \camone{under} (a) a low-contention \camone{and synchronization-intensive scenario} using 4 NDP units, and (b) a high-contention \camone{scenario} using 2 and 4 NDP units.}
  \label{fig:hier_benefit}
  \vspace{-3pt}
\end{figure}

\noindent\textbf{High-contention.} 
\camfour{Figure~\ref{fig:hier_benefit}b evaluates this scenario using a queue data structure with four different \camthree{link} latency values \camthree{between} NDP units, for 30 and 60 NDP cores. \myName{}} \camthree{with 30 NDP cores outperforms \emph{flat} from 1.23$\times$ to 1.76$\times$, \camfour{as} the inter-NDP-unit latency \camfour{increases from 40 ns to 500 ns (i.e., with increasing non-uniformity in the system).}} \camfour{In a scenario with high non-uniformity in the system and large number of contended cores,
e.g., using a 500 \camthree{ns} inter-NDP-unit latency and 60 NDP cores,} \camthree{\myName{}'s benefit increases \camtracy{to a} 2.14$\times$ speedup over \emph{flat}}. We \camfour{conclude that} \myName{} performs \emph{significantly} better than \emph{flat} \camfour{under high-contention}.


\vspace{1pt}

Overall, \camtwo{we conclude that in \emph{non-uniform}, \emph{distributed} NDP systems, \emph{only} a \emph{hierarchical} hardware synchronization design can achieve high performance under \emph{all} various scenarios.}

} 
\vspace{1pt}

\subsubsection{\myTableShort{} \camone{S}ize}
\begin{figure}[H]
  \centering
  \includegraphics[scale=0.026]{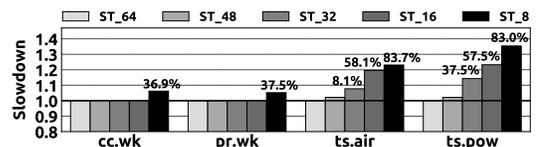}
  \caption{\camtwo{Slowdown with varying} \myTableShort{} size \camtwo{(normalized to 64-entry \myTableShort{})}. \camone{Numbers on top of bars show} the percentage of overflowed requests.}
  \label{fig:sens_table}
\end{figure}

We show the effectiveness of the proposed 64-entry \myTableShort{} \camtwo{(per NDP unit)} using real applications. Table~\ref{tab:graphs_occ} shows \camtwo{the measured} occupancy across all \myTableShort{}s. Figure~\ref{fig:sens_table} shows the \camone{performance sensitivity} to \myTableShort{} size. In graph applications, the average \myTableShort{} occupancy is low (2.8\%), and the 64-entry \myTableShort{} never overflows: \camtwo{maximum} occupancy \camtwo{is} 63\% (\emph{cc.wk}). 
\camtwo{In contrast, time series analysis has higher \myTableShort{} occupancy (reaching up to 89\% in \emph{ts.pow}) due to the high synchronization intensity, but there are no \myTableShort{} overflows. Even a 48-entry \myTableShort{} overflows for only 0.01\% of synchronization requests, and incurs 2.1\% slowdown over a 64-entry \myTableShort{}.}
We conclude that the proposed 64-entry \myTableShort{} meets the needs of applications that have high synchronization intensity.


\begin{center} 
\resizebox{0.976\columnwidth}{!}{%
 \begin{tabular}{l c S[table-format = 2.2]} 
 \toprule
 \textbf{\myTableShort{} Occupancy} & \textbf{Max (\%)} & \textbf{Avg (\%)} \\ 
 \midrule
 \midrule
  \textbf{bfs.wk} & 51 & 1.33 \\
 \textbf{bfs.sl} & 59 & 1.49 \\
 \textbf{bfs.sx} & 51 & 3.24 \\
 \textbf{bfs.co} & 55 & 6.09 \\
 \textbf{cc.wk} & 63 & 1.27 \\
 \textbf{cc.sl} & 61 & 2.16 \\
 \textbf{cc.sx} & 48 & 2.43 \\
 \textbf{cc.co} & 46 & 4.53 \\
 \textbf{sssp.wk} & 62 & 1.18 \\
 \textbf{sssp.sl} & 54 & 2.08 \\
 \textbf{sssp.sx} & 50 & 2.20 \\
 \textbf{sssp.co} & 48 & 5.23\\
 \textbf{pr.wk} & 62 & 4.27 \\
 \bottomrule
 \end{tabular}
 ~ \hspace{18pt}
 \begin{tabular}{l c S[table-format = 2.2]} 
 \toprule
 \textbf{\myTableShort{} Occupancy} & \textbf{Max (\%)} & \textbf{Avg (\%)} \\ 
 \midrule
 \midrule
 \textbf{pr.sl} & 51 & 2.27 \\
 \textbf{pr.sx} & 53 & 2.46 \\
 \textbf{pr.co} & 48 & 4.72 \\
 \textbf{tf.wk} & 62 & 1.44 \\
 \textbf{tf.sl} & 53 & 2.21 \\
 \textbf{tf.sx} & 50 & 2.99 \\
 \textbf{tf.co} & 48 & 4.61 \\
 \textbf{tc.wk} & 62 & 1.26 \\
 \textbf{tc.sl} & 48 & 2.08 \\
 \textbf{tc.sx} & 50 & 2.77 \\
 \textbf{tc.co} & 51 & 4.52 \\
 \hline 
 \textbf{ts.air} & 84 & 44.20 \\
 \textbf{ts.pow} & 89 & 43.51 \\
 \bottomrule
 \end{tabular}} 
 \captionof{table}{\label{tab:graphs_occ}\myTableShort{} occupancy in real applications.}
\end{center}

\vspace{-3pt}

\subsubsection{Overflow Management}\label{OverflowEvalbl}
The linked list and BST\_FG data structures are the \emph{only} cases where the proposed 64-entry \myTableShort{} overflows, when using 60 cores, \camone{for} 3.1\% and 30.5\% \camone{of the requests}, respectively. This is because each core requests at least two locks \emph{at the same time} \camone{during the execution}. Note that these \camtwo{synthetic} benchmarks \camone{represent} extreme scenarios, where all cores repeatedly perform key-value operations.

\begin{figure}[H]
  \centering
  \includegraphics[scale=0.26]{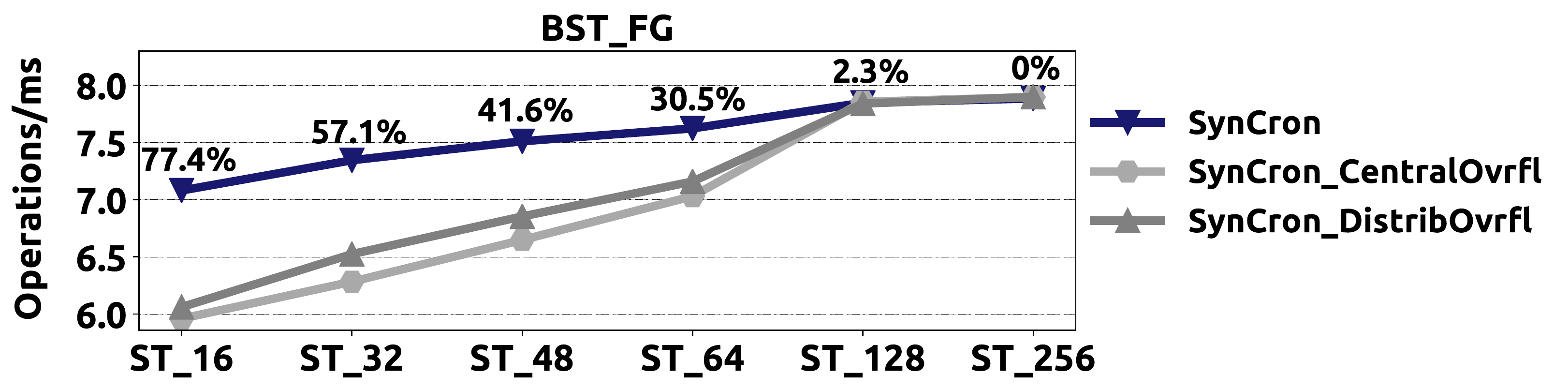}
  \caption{Throughput achieved by BST\_FG using different overflow schemes and varying the \myTableShort{} size. The reported numbers \camtwo{show} to the percentage of overflowed requests.}
  \label{fig:overflow}
\end{figure}

Figure~\ref{fig:overflow} compares \camone{BST\_FG's performance with} \myName{}'s integrated overflow scheme \camone{versus} with a non-integrated scheme as in MiSAR. \camone{When overflow occurs, MiSAR's accelerator aborts all participating cores notifying them to use an alternative synchronization library, \camtwo{and when the cores finish synchronizing
via an alternative solution, they notify MiSAR’s accelerator to switch back to hardware synchronization. We adapt this scheme to \myName{} for comparison purposes:} when an \myTableShort{} overflows, \myEngineShort{}s send abort messages to NDP cores with \camthree{a} hierarchical protocol, notifying them to use an alternative synchronization solution, \camtwo{and after finishing synchronization they notify \myEngineShort{}s to decrease their indexing counters and switch to hardware.}} We evaluate two alternative solutions: (i) \emph{SynCron\_CentralOvrfl}, where one dedicated NDP core handles all synchronization variables, and (ii) \emph{SynCron\_DistribOvrfl}, where \camone{one} NDP core per NDP unit handles variables located in the same NDP unit. With 30.5\% overflowed \camone{requests} (i.e., with a 64-entry \myTableShort{}),
\camtwo{\emph{SynCron\_CentralOvrfl} and \emph{SynCron\_DistribOvrfl} incur 12.3\% and 10.4\% performance slowdown \camthree{compared to} with \emph{no} \myTableShort{} overflow, due to high network traffic and communication costs between NDP cores and \myEngineShort{}s.} \camone{In contrast},
\camtwo{\myName{} affects performance \camthree{by only} 3.2\% \camthree{compared to} with \emph{no} \myTableShort{} overflow.}
We conclude that \myName{}'s integrated hardware-only overflow scheme \camtwo{enables} \camone{very small performance overhead.}

\subsection{\myName{}'s Area and \camfour{Power} Overhead} \label{Areabl}

Table~\ref{tab:area} compares \camtwo{an} \myEngineShort{} with \camtwo{the} ARM Cortex A7 core~\cite{ARM_CORTEX}. \camone{
\camtwo{We estimate the SPU using Aladdin~\cite{shao2016aladdin}, and the \myTableShort{} and indexing counters using CACTI~\cite{Muralimanohar2007Optimizing}.}
We conclude that \camnine{our proposed hardware unit} incurs very modest area and \cameight{power} costs to be integrated \camtwo{into} the compute die of an NDP unit.}

\begin{center} 
\resizebox{0.9\columnwidth}{!}{%
 \begin{tabular}{l c c} 
 \toprule
 \textbf{} & \textbf{\myEngineShort{} \camtwo{(Synchronization Engine)}} & \textbf{ARM Cortex A7~\cite{ARM_CORTEX}} \\ 
 \midrule
 \midrule
 \textbf{Technology} & 40nm & 28nm \\
 \hline
 \multirow{3}{*}{\shortstack[l]{\textbf{Area}}} & SPU: 0.0141mm\textsuperscript{2}, \myTableShort{}: 0.0112mm\textsuperscript{2} & \multirow{2}{*}{\shortstack[l]{32KB L1 Cache}} \\
  & Indexing Counters: 0.0208mm\textsuperscript{2} & \\
  & \textbf{Total:} 0.0461mm\textsuperscript{2} & \textbf{Total:} 0.45mm\textsuperscript{2} \\
 \hline
 \textbf{Power} & 2.7 mW & ~100mW \\
 \bottomrule
 \end{tabular}
 } 
 \captionof{table}{\label{tab:area}\camone{Comparison of \myEngineShort{} with a simple general-purpose in-order core, \camtwo{ARM Cortex A7}}.}
 \vspace{2pt}
\end{center}

\section{Related Work}
\label{Relatedbl}


To our knowledge, \camtwo{our work is the first one to (i) comprehensively analyze and evaluate synchronization primitives in NDP systems, and (ii) propose an end-to-end hardware-based synchronization mechanism for efficient execution of such primitives. We briefly discuss prior work.}

\textbf{Synchronization on NDP.} 
Ahn et al.~\cite{Ahn2015Scalable} include a \mpsync{} barrier similar to our \naive{} \camtwo{baseline}. Gao et al.~\cite{Gao2015Practical} implement a \camtwo{hierarchical} tree-based barrier for HMC~\cite{HMC}, where cores first synchronize inside the vault, then across vaults, and finally across HMC \camone{stacks}. Section~\ref{Performancebl} shows that \myName{} outperforms such schemes. Gao et al.~\cite{Gao2015Practical} also provide remote atomics at the vault controllers of HMC. However, synchronization \camtwo{using} remote atomics creates high global traffic and hotspots~\cite{Wang2019Fast,Mukkara2019PHI,li2015fine,yilmazer2013hql,eltantawy2018warp}.

\textbf{Synchronization on CPUs.}
A range of hardware synchronization mechanisms have been proposed for commodity CPU systems~\cite{abell2011glocks,sampson2006exploiting, abellan2010g, oh2011tlsync,Sergi2016WiSync,akgul2001system}. These are not suitable for NDP systems because they either (i) rely on the underlying cache coherence system~\cite{sampson2006exploiting,akgul2001system}, (ii) are tailored for \camtwo{the} 2D-mesh network topology \camone{to connect} all cores~\cite{abell2011glocks,abellan2010g}, or (iii) use transmission-line technology~\cite{oh2011tlsync} or on-chip wireless technology~\cite{Sergi2016WiSync}. 
Callbacks~\cite{Ros2015Callback} includes a directory cache structure close to the LLC of a CPU system \camsix{built on} self-invalidation coherence protocols~\cite{Kaxiras2010SARC,Choi2011DeNovo,Kaxiras2013ANew,Sung2014DeNovoND,Lebeck1995Dynamic,Ros2012Complexity}. Although it has low area cost, it would be oblivious to the non-uniformity of NDP, thereby incurring high performance overheads \camone{\camtwo{under} high} contention (Section~\ref{Flat}). \camone{Callbacks 
improves performance of spin-wait \camtwo{in hardware}}, 
on top of which \camtwo{high-level primitives (locks/barriers) are implemented in software.} In contrast, \myName{} directly supports high-level primitives \camtwo{in hardware}, and is tailored \camthree{to} all \camone{salient} characteristics of NDP systems.

The closest works to ours are SSB~\cite{Zhu2007SSB}, LCU~\cite{Vallejo2010Architectural}, and MiSAR~\cite{Liang2015MISAR}. SSB, a shared memory scheme, includes a small buffer attached to each controller of LLC to provide lock semantics for a given data address. LCU, a \mpsync{} scheme, incorporates a control unit \camtwo{into} each core and a reservation table \camtwo{into} each memory controller to provide reader-writer locks. MiSAR is a \mpsync{} synchronization accelerator distributed at each LLC slice of tile-based many-core chips. \camone{These schemes provide efficient synchronization for CPU systems \camtwo{\emph{without}} relying on hardware coherence protocols. \camtwo{As shown in Table~\ref{tab:comparison}, compared to these works}, \myName{} is a more effective, general and easy-to-use solution for NDP systems. These works have two major shortcomings. First, they are designed for \emph{uniform} architectures, and would incur \camtwo{high} performance overheads in \emph{non-uniform, distributed} NDP systems under high-contetion scenarios, similarly to \emph{flat} \camtwo{in  Figure~\ref{fig:hier_benefit}b.} Second, SSB and LCU handle overflow cases using software exception handlers that typically incur large performance overheads, while MiSAR's overflow \camtwo{scheme} would incur high performance degradation due to high network traffic and communication costs between the cores and the synchronization accelerator \camtwo{(Section~\ref{OverflowEvalbl})}. In contrast, \myName{} is a \camtwo{non-uniformity} aware, hardware-only, end-to-end solution designed to \camtwo{handle key} characteristics of NDP systems.}

\textbf{Synchronization on GPUs.} 
GPUs support remote atomic units at the shared cache and hardware barriers among threads of the same block~\cite{teslav100}, while inter-block barrier synchronization is inefficiently implemented via the host CPU~\cite{teslav100}. The closest work to ours is HQL~\cite{yilmazer2013hql}, which modifies the tag arrays of L1 and L2 caches to support \camone{the} lock primitive. This scheme incurs high area \camnine{cost}~\cite{eltantawy2018warp}, and is tailored to the GPU architecture that includes a \camtwo{shared} L2 cache, while most NDP systems do \emph{not} have shared \camtwo{caches}.

\textbf{Synchronization on MPPs.}
The Cray T3D/T3E~\cite{kessler1993crayTA,scott1996synchronization}, SGI Origin~\cite{Laudon1997SGI}, and AMOs~\cite{zhang2004highly} include remote atomics at the memory controller, while NYU \camone{Ultracomputer}~\cite{gottlieb1998NYU} \camtwo{provides} \emph{fetch\&and} remote atomics in each network switch. As discussed in Section~\ref{Motivationbl}, synchronization via remote atomics \camone{incurs} high performance overheads due to high global traffic~\cite{Wang2019Fast,Mukkara2019PHI,yilmazer2013hql,eltantawy2018warp}. Cray T3E supports a barrier \camone{using} physical wires, \camtwo{but it is} designed \camtwo{specifically} for 3D torus interconnect. Tera \camtwo{MTA}~\cite{Alverson1990Tera}, HEP~\cite{Jordan1983Performance,Smith1978Pipelined}, J- and M-machines~\cite{Dally1992TheMessage,Keckler1998Exploiting}, and Alewife~\cite{Agarwal1995Alewife} provide synchronization using hardware bits (\emph{full/empty} bits) as tags in \emph{each memory word}. This \camtwo{scheme can incur} high area cost~\cite{Vallejo2010Architectural}. QOLB~\cite{Goodman1989EfficientSP} associates one cache line for every lock to track a pointer to the next waiting core, and one cache line for local spinning using bits (\emph{syncbits}). QOLB is \camtwo{built on} the underlying cache coherence protocol. Similarly, DASH~\cite{Lenoski1992Dash} keeps a queue of waiting cores for a lock in the directory used for coherence to notify caches when the lock is released. CM5~\cite{Leiserson1992CM5} supports remote atomics and a barrier among cores via a dedicated physical control network (organized as a binary tree), which would incur high \camnine{hardware cost to be supported in NDP systems.}

\section{Conclusion}
\label{Conclusionbl}
\setstretch{0.84}
\myName{} is \camone{the first} end-to-end synchronization \camfour{solution} for NDP systems. \myName{} avoids the need for complex coherence protocols and expensive \emph{rmw} operations, incurs very modest hardware cost, \camone{generally supports many synchronization primitives} and is easy-to-use. Our evaluations show that it outperforms prior designs under various conditions, providing high performance both \camtwo{under} \camone{high-contention} (due to \camone{reduction of} expensive traffic across NDP units) and low-contention scenarios (due to direct buffering \camone{of synchronization variables} and \camone{high} execution parallelism). We conclude that \myName{} is an \camthree{efficient} synchronization \camfour{mechanism} for NDP systems, and hope that this work encourages \camone{further} \camtwo{comprehensive studies} of the synchronization problem in heterogeneous systems, including NDP systems.

\setstretch{0.824}
\section*{Acknowledgments}
\vspace{-1pt}
We thank the anonymous reviewers of ISCA 2020, MICRO 2020 and HPCA 2021 for feedback. We thank Dionisios Pnevmatikatos, Konstantinos Nikas, Athena Elafrou, Foteini Strati, Dimitrios Siakavaras, Thomas Lagos, Andreas Triantafyllos for helpful technical discussions. We acknowledge support from the SAFARI group's industrial partners, especially ASML, Google, Facebook, Huawei, Intel, Microsoft, VMware, and Semiconductor Research Corporation. During part of this research, Christina Giannoula was funded 
from the General Secretariat for Research and Technology (GSRT) and the Hellenic Foundation for Research and Innovation (HFRI).

\vspace{-1pt}

%

\setstretch{0.83}
\SetTracking
 [ no ligatures = {f},
 outer kerning = {*,*} ]
 { encoding = * }
 { -40 } 

{

  \let\OLDthebibliography\thebibliography
  \renewcommand\thebibliography[1]{
    \OLDthebibliography{#1}
    \setlength{\parskip}{0pt}
    \setlength{\itemsep}{0pt}
  }
  \bibliographystyle{IEEEtranS}
  \bibliography{ref}
}

\end{document}